\documentclass[american,english,notitlepage]{revtex4-1}
\usepackage[T1]{fontenc}
\usepackage[latin9]{inputenc}
\usepackage{wrapfig}
\usepackage{amsmath}
\usepackage{amsthm}
\usepackage{amssymb}
\usepackage{graphicx}

\makeatletter

\usepackage[usenames,dvipsnames]{pstricks}
\usepackage{epsfig}
\usepackage{pst-grad} 
\usepackage{pst-plot} 
\usepackage[space]{grffile} 
\usepackage{etoolbox} 
\usepackage{hyperref}

\makeatother

\usepackage{babel}
\begin{document}

\title{Lower-Hybrid Drift Instability and Macroscopic Flow of Colliding
Magnetized Plasmas}

\author{M.A. Malkov$^{1}$, V.I. Sotnikov$^{2}$}

\affiliation{$^{1}$CASS and Department of Physics, University of California,
San Diego, La Jolla, CA 92093\\$^{2}$Air Force Research Laboratory,
Sensors Directorate, WPAFB, OH 45433}
\begin{abstract}
Microscopic instability and macroscopic flow pattern resulting from
colliding plasmas are studied analytically in support of laboratory
experiments. The plasma flows are assumed to stream radially from
two separate centers. In a quasi-planar (2D) geometry, they may arise
from an Ohmic explosion of two parallel wires, but similar configurations
emerge from other outflows, e.g., colliding winds in binary star systems.
One objective of this paper is to characterize the flow instabilities
developing near the flow stagnation line. An exact solution for the
Buneman-type dispersion equation is obtained without conventional
simplifications. The unstable wave characteristics are key to anomalous
resistivity that determines the reconnection rate of opposite magnetic
fields transported with each flow toward the stagnation zone. The
second objective of the paper is to calculate the stream function
of the plasma shocked upon collision. We addressed this task by mapping
the flow region to a hodograph plane and solving a Dirichlet problem
for the stream function. By providing the instability growth rate,
responsible for anomalous transport coefficients, and the overall
flow configuration, these studies lay the ground for the next step.
From there, we will examine the field reconnection scenarios and emerging
mesoscopic structures, such as radial striata observed in the experiments.

\section{Introduction}

\end{abstract}
\maketitle
\begin{wrapfigure}{o}{0.5\columnwidth}%
\includegraphics[bb=0bp 280bp 612bp 720bp,scale=0.4]{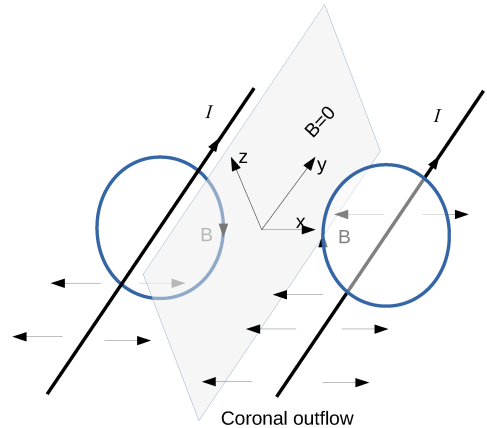}

\caption{Schematics of the experimental setup. Two wires, carrying current
$I$ each, are directed along the $y-$ axis and connect the chamber
electrodes (not shown). \label{fig:Schematics-of-the}}
\end{wrapfigure}%

\begin{wrapfigure}{o}{0.5\columnwidth}%
\includegraphics[bb=0bp 300bp 612bp 740bp,scale=0.3]{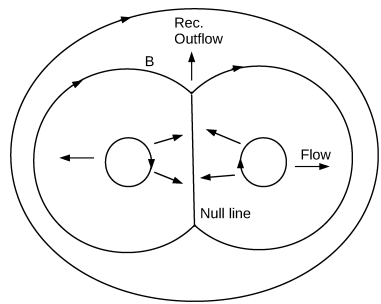}

\caption{Reconnection flow initiated by parallel (out of plane) currents.\label{fig:Reconnection-flow-initiated} }
\end{wrapfigure}

Colliding plasma flows with nearly opposite magnetic field orientation,
often supersonic and superalfvenic, is an exciting and challenging
problem to study. It has numerous applications in laboratory and space
plasmas. The list of relevant systems includes, but is not limited
to, colliding winds from two neighboring stars (often called binaries),
corotating interaction regions in a solar wind plasma near the ecliptic
plane, flows associated with the coronal mass ejection from the Sun,
or even wall reflected counterpropagating plasma stream in hybrid
and PIC simulations of collisionless shocks. In laboratory plasmas,
one of the simplest configurations leading to colliding plasma flows
can be created by driving strong unidirectional currents through a
pair of parallel wires. The azimuthal magnetic fields generated around
each wire, and the Ohmic current dissipation and heating occurring
upon wire evaporation, launches strong radial outflows of magnetized
plasmas. Upon colliding with each other, they form a flow pattern
highly suggestive of magnetic field reconnection, and the development
of various plasma instabilities, Figs. \ref{fig:Schematics-of-the}
and \ref{fig:Reconnection-flow-initiated}. Indeed, the symmetry plane
between the wires must be the place where magnetic field lines from
the opposing flows reconnect, Fig.\ref{fig:Reconnection-flow-initiated},
as they need to conform to a single circular field configuration at
distances much larger than the gap between the wires. 

The reconnection efficiency strongly depends on the flow dissipation
rate in the reconnection zone. In the case of hot outflows not significantly
cooled by expansion, or even additionally heated upon their interaction
with each other, the plasma binary collisions remain rare. In this
case the anomalous resistivity, supported by plasma micro-instabilities
that are likely to develop in the flows, takes the role of the classical
resistivity. The potential of the flows to drive strong instabilities
is the first topic of this paper. Knowing their growth rates and how
are they distributed in the flow collision zone will allow one to
understand the overall flow organization. These two sides of the problem
are strongly related and should be treated on an equal footing which
is, however, a formidable task. Therefore, our strategy consists of
determining the instability growth rate making only most general assumptions
about the flow and wave parameters. In particular, selecting the lower-hybrid
drift instability (LHDI) as the most potent one to support the anomalous
transport, we calculate its local growth rate with no further approximations.
In other words, we solve the well-known dispersion relation for the
LHDI (or any other two-stream type instability for that matter) exactly.
In certain plasma flows, however, particularly some of the laser-evaporated
target flows, there is no much room or time for the development of
plasma micro-instabilities. With this regard, we refer the reader
to the paper \cite{Ryutov2013PhPlreconnection}, where such colliding
plasma flows with strong intraflow collisions have been studied in
detail. 

The second topic of the paper concerns the flow pattern emerging upon
collision of outflows from two parallel wires. As in the instability
part of our study, our strategy here is to calculate the stream function
of the flow making only the most general assumptions about its character.
One such assumption is the \emph{insignificant }dynamic role of the
magnetic field. This simplification, supported by experiments \cite{Sarkisov2005},
allows us to calculate the stream function of colliding flows between
two shocks formed upon collision using gas-dynamic rather than magnetohydrodynamic
equations. As our research of the both topics is performed within
a fairly general framework, we are planning to combine them to describe
the flow collision phenomenon, followed by the magnetic energy dissipation through
reconnection supported by the anomalous resistivity. Again, in the
paper \cite{Ryutov2013PhPlreconnection}, an alternative situation
of colliding MHD flows has been addressed with a primary emphasis
on the magnetic aspect of the flow.

The paper is organized according to the following sections. Sec.\ref{sec:A-Model-for}
discusses the overall properties of the colliding flows. It further
deals with the derivation of the dispersion relation for the LHDI
and its closed form solution. Sec.\ref{sec:Macroscopic-Properties-of}
presents a more elaborate treatment of the post-shock flow based on
a hodograph map of its area and derivation of the Chaplygin equations.
The general solution for the stream function is presented. The equations
for the parameters of the maps are given in Appendices. For illustration,
a simplified flow example is also given. The paper concludes with
a brief discussion of its topics, the summary main results, and an
outline of the next steps.

\section{A Model for Colliding Plasma Flows\label{sec:A-Model-for}}

\subsection{Sketch Description of the Flow\label{subsec:Sketch-Description-of}}

Before the two plasma flows reach a magnetic null ($B\approx0$),
where they collide, partly interpenetrate and spread sideways, each
of them undergoes a substantial evolution. This part of the problem
has already been considered by others using a single wire setup (see,
e.g., \cite{Sarkisov2005} and references therein). Many efforts have
been made in conjunction with $z-$ pinches. According to a universally
accepted picture, the wire explosion can be broken down roughly into
the following two phases. 

Initially, most of the current between the electrodes flows through
the wire that undergoes a complicated process of melting and vaporization.
The details of these phenomena are, however, not so important for
the present study. Next, when the plasma is created from the evaporated
wire material and heated significantly, it forms an expanding \emph{corona}.
During this process, the current jumps from the wire to the corona.
Some important physical phenomena occurring during the corona expansion,
have been discussed and simulated, primarily within various MHD models
\cite{Chittenden2001,Sarkisov2005}. However, it has also been realized
that some microphysics not captured by the MHD approach is crucial
to the macroscopic behavior of the corona. In particular, as the corona
is rarefied and hot, it can transition into a collisionless regime.
In this case, the macroscopic transport properties must be anomalous. 

Turning to the double-wire configuration, there is no shortage of
free energy sources for various instabilities that may play their
parts in momentum and energy exchange between particles in each flow
and between the colliding plasmas as a whole. So, the classical Spitzer
resistivity needs to be complemented, if not replaced, by the anomalous
one. A notably efficient instability, which in many cases has the
lowest threshold, is the lower-hybrid drift instability (LHDI). While
this instability has already been argued \cite{Chittenden1995} to
play an important role in a single-wire plasma expansion, we will
study it in an environment created by two colliding coronal plasmas
in the vicinity of the neutral plane. We will also significantly generalize
the existing results by solving the LHDI dispersion equation exactly. 

\subsection{General Dispersion Equation for the LHDI Instability\label{subsec:General-Dispersion-Equation}}

Strictly speaking, the system under consideration cannot be subjected
to the conventional stability analysis starting from an equilibrium
plasma state since the flow is time-dependent. However, as the LHDI
is usually very fast, we may assume that the flow remains quasi-stationary
over the instability e-folding time. Turning to the spatial scales
of the problem we note that plasma is essentially inhomogeneous in
at least two dimensions perpendicular to the wire direction. At the
same time, the maximum growth rate is reached at scales comparable
to, or even slightly smaller than, the electron gyroradius, $\rho_{e}$.
The latter is, in turn, much smaller than all relevant scales (except
the Debye length) associated with the plasma and magnetic field inhomogeneity.
Therefore, we carry out the stability analysis based on a local approximation.
We will use a coordinate system with the $x-$ axis running normally
to the wires in their plane, $y$ - parallel to the wires and $z$
will then be the vertical off-plane axis, Fig.\ref{fig:Schematics-of-the}. 

Based on the above considerations, the ion gyroradius strongly exceeds
the characteristic wavelength of the instability, and the ion gyrofrequency
is also much smaller than the growth rate. We will, therefore, treat
ions as unmagnetized. Because of the electric field $E_{y}$, we include
the ion drift velocity component in $y-$ direction, $V_{di}\sim\sqrt{eE_{y}y/M}.$
$E_{y}$ is, however, not the only electric field component present
in the plasma near the neutral plane. Due to effects of charge separation
in colliding plasmas, associated primarily with the different gyroradii
of electrons and ions, the $E_{x}$ component must be quite strong.
In addition, the two colliding plasmas may partially interpenetrate
which should also build up electric field, $E_{x}$. More importantly,
it will lead to an electron $E\times B$ drift in the direction of
wave propagation, that is, $V_{Ey}\sim cE_{x}/B$. We neglect the
electron motion in $z$ direction, assuming that $k_{z}V_{ze}\ll k_{y}V_{ye},\;\omega$.
It is worth noting, however, that for sufficiently large $k_{z}$,
the phase velocity of oscillations in $z-$ direction, $\left(\omega-k_{y}V_{ye}\right)/k_{z}$,
may approach the electron thermal velocity which would result in a
fast electron acceleration along the magnetic field. The field is
oriented parallel to $z-$ axis in the plane of the wires, but it
is globally circular as it originates from the currents flowing through
the wires. 

Using a local approximation, we test stability of the following perturbation,
$\propto\exp\left(-i\omega t+ik_{y}y+ik_{z}z\right)$, assuming also
that $k_{z}\ll k_{y}$. The x-structure of the mode is ignored, as
usual in the local approximation. In $y$- and $z$-directions, the
plasma is considered to be homogeneous, even though the ion drift
velocity, for example, slowly depends on y. Regarding the $z$-dependence,
this approximation is valid not far away from the plane $z=0$. We
assume of course that $L_{y}k_{y}\gg1$, where $L_{y}$ is the size
of the chamber in wire direction. Also, the presence of ion-neutral
collisions should bring the ion drift velocity in the electric field
direction to equilibrium, so the approximation of constant velocity
appears plausible for the short-wave instability. There are also electron
and ion drifts and currents, associated with the magnetic field, temperature
and density variations in the $x$-direction that in general need
to be included in the electron and ion responses to the wave perturbation.

A sufficiently general dispersion equation for the LHDI in electrostatic
approximation, after obvious modifications suggested by our discussion
above, can be adopted from, e.g. Ref. \cite{DavidsonGladd1975}:

\begin{equation}
1+\frac{2\omega_{pi}^{2}}{k_{y}^{2}V_{Ti}^{2}}\left[1+\xi Z\left(\xi\right)\right]+\frac{\omega_{pe}^{2}}{\omega_{ce}^{2}}\frac{1-I_{0}\left(b\right)e^{-b}}{b}-\frac{k_{z}^{2}}{k_{y}^{2}}\frac{\omega_{pe}^{2}}{\left(\omega-k_{y}V_{ey}\right)^{2}}+\frac{2\omega_{pe}^{2}}{k_{y}^{2}V_{Te}^{2}}\frac{k_{y}V_{\Delta}}{\omega-k_{y}V_{ey}}=0\label{eq:DispEqGen}
\end{equation}
Here we have used the following notations,

\[
\xi=\frac{\omega-k_{y}V_{di}}{\left|k_{y}\right|V_{Ti}},\;\;\;Z\left(\xi\right)=-e^{-\xi^{2}}\left(2\int_{0}^{\xi}e^{z^{2}}dz-i\sqrt{\pi}\right),\;\;\;b=\frac{\left|k_{y}\right|V_{Te}}{\omega_{ce}}
\]

\[
V_{\Delta}=-\frac{V_{Te}^{2}}{\omega_{ce}}I_{0}\left(b\right)e^{-b}\left\{ \kappa_{n}-\kappa_{B}\left[1-b\left(1-I_{1}/I_{0}\right)\right]-\kappa_{T}\left(1-I_{1}/I_{0}\right)\right\} ,
\]
where $I_{0,1}$ are the modified Bessel functions, $\kappa$'s are
the inverse scales of respective inhomogeneity, $\kappa_{n}=n^{-1}dn/dx$,
$\kappa_{B}=B^{-1}dB/dx$, and $\kappa_{T}=T_{e}^{-1}dT_{e}/dx$.

As we emphasized, in a flow resulting from the collision of two expanding
plasmas the main instability driver is likely to be a relative motion
of electrons and ions in the direction of wave propagation, that is,
along the $y-$ axis. Therefore, to determine the most favorable conditions
for the instability, we may neglect the contributions of particle
drifts associated with the inhomogeneity, collected in the term $V_{\Delta}$.
We will also simplify the dispersion equation somewhat further using
the following dimensionless variables

\[
\hat{\omega}=\frac{\omega}{\left|k_{y}\right|V_{T_{i}}},\;\;\;\eta=\frac{\omega_{pi}^{2}}{k_{y}^{2}V_{Ti}^{2}},\;\;\;\Omega_{i}=\frac{V_{di}}{V_{Ti}}\frac{k_{y}}{\left|k_{y}\right|},\;\;\;\Omega_{e}=\frac{V_{ey}}{V_{Ti}}\frac{k_{y}}{\left|k_{y}\right|}
\]
Since we expect the relative bulk motion of electrons and ions to
be the main driver of the instability, we neglect the thermal ion
spread as well. The latter simplification translates into the condition$\left|\xi\right|\gg1$
and the dispersion equation rewrites

\begin{equation}
\mathcal{F}\left(\hat{\omega}\right)\equiv\frac{1}{\left(\hat{\omega}-\Omega_{i}\right)^{2}}+\frac{K}{\left(\hat{\omega}-\Omega_{e}\right)^{2}}=\frac{1}{\eta}+\frac{T_{i}}{T_{e}}b\left(1-I_{0}\left(b\right)e^{-b}\right)\equiv A\label{eq:DispEq2}
\end{equation}
where 

\begin{equation}
K=\left(M/m\right)\left(k_{z}/k_{y}\right)^{2}\label{eq:Kdef}
\end{equation}

Thus, we arrived at a fourth order algebraic equation for the frequency
$\hat{\omega}$. Apart from the notation, the same equation is widely
used in studies of Buneman and modified two-stream instabilities,
e.g. \cite{LiewerKrall1973,DavidsonGladd1975}. Of course, it can
be solved exactly for $\hat{\omega}$ using one of the solutions obtained
by Ferrari, Euler and some other famous mathematicians of the past.
At the same time, these solutions are quite laborious and impractical
to use. ``Mathematica,'' for example, returns a set of four solutions
filling the entire computer screen. Not surprisingly, equations similar
to eq.(\ref{eq:DispEq2}) are usually solved either approximately
or numerically. The numerical solutions will not be helpful in analytic
calculations of the anomalous transport coefficients that are crucial
for obtaining the macroscopic flow. The approximate solutions, on
the other hand, are highly restrictive in situations where the flow
is not known beforehand. 

Indeed, our goal is to characterize the instability for the flow conditions
changing arbitrarily in time and space. Also, a macroscopic back reaction
of the developing instability on the flow itself will result in such
changes. Therefore, it is highly desirable to obtain an exact solution
to eq.(\ref{eq:DispEq2}) that would cover the entire parameter space
without restrictions. We tackle this problem in the next subsection
and demonstrate that such solution can, in fact, be quite manageable,
by contrast to the full algebraic solution of the quartic eq.(\ref{eq:DispEq2}).
The novelty of our approach is in that we separate the two complex
roots from the two real roots, thus simplifying the solution tremendously. 

\subsection{Instability Analysis\label{subsec:Instability-Analysis}}

Let us start with summarizing simple and well-known aspects of eq.(\ref{eq:DispEq2}).
First, as $A>0,$ two of the four solutions are purely real. They
have the maximum and minimum real parts among the four solutions.
The remaining two roots, which are between the poles at $\hat{\omega}=\Omega_{e,i}$,
are also real if $A-\mathcal{F}_{{\rm min}}>0$. The interesting case
is when they are complex conjugate, that is when $A-\mathcal{F}_{{\rm min}}\left(\hat{\omega}\right)<0$.
One of the roots corresponds then to a positive growth rate $\hat{\gamma}=\Im\hat{\omega}$.
This unstable root can be easily found near the instability threshold:

\[
\Im\hat{\omega}\equiv\hat{\gamma}=\frac{4\delta^{2}K^{1/6}}{\sqrt{3}\left(1+K^{1/3}\right)^{5/2}}\sqrt{\mathcal{F}\left(\hat{\omega}_{0}\right)-A}
\]
Here $\delta=\left(\Omega_{e}-\Omega_{i}\right)/2$ (assumed positive),
and $\hat{\omega}_{0}$ is where $\mathcal{F}$ has a minimum, $\Omega_{i}<\hat{\omega}_{0}<\Omega_{e}$.
There are a few other limiting cases of eq.(\ref{eq:DispEq2}) that
are frequently considered. For $K=1$, one obtains a simple exact
result when the equation can be transformed to a biquadratic form.
In other cases, such as large or small $K$, the equation is treated
only approximately.

Our goal, however, is to obtain exact expressions for both the frequency
and the growth rate of unstable waves depending on parameters. As
these are coordinate dependent, this will locate the wave generation
domain, and therefore that of the current and flow dissipation. To
this end, we simplify eq.(\ref{eq:DispEq2}) by reducing the number
of parameters it depends on, from four to two. By introducing a new
dimensionless frequency 

\begin{equation}
\Omega=\hat{\frac{\omega}{\delta}}-\frac{1}{2\delta}\left(\Omega_{i}+\Omega_{e}\right)\label{eq:OmegaDef}
\end{equation}
and denoting $B=\delta^{2}A$, eq.(\ref{eq:DispEq2}) rewrites

\begin{equation}
\frac{1}{\left(\Omega+1\right)^{2}}+\frac{K}{\left(\Omega-1\right)^{2}}=B\label{eq:DispEq3}
\end{equation}
The parameter $B$ depends on the dimensionless wave number $b$ as
follows

\begin{equation}
B=\frac{T_{i}}{T_{e}}\delta^{2}b^{2}\left[\frac{\omega_{ce}^{2}}{\omega_{pe}^{2}}+\frac{1-I_{0}\left(b\right)e^{-b}}{b}\right]\label{eq:Bdef}
\end{equation}
We note that for small $b$, the second term in the brackets behaves
as $1-3b/4$. 

\begin{wrapfigure}{o}{0.5\columnwidth}%
\includegraphics[bb=0bp 0bp 360bp 210bp,scale=0.5]{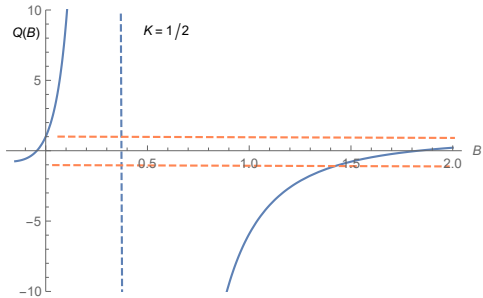}

\caption{The function $Q\left(B,K\right)$ on the r.h.s. of eq.(\ref{eq:zeta})
shown at $K=1/2$. The vertical dashed line indicates the singularity
at $B=\left(K+1\right)/4$. The line $Q=-1$ intersects $Q\left(B\right)$
at $B=B_{0}$, so that the condition $\left|Q\left(B\right)\right|<1$
is met only for $B>B_{0}$ where no complex roots are found. \label{fig:QofB}}
\end{wrapfigure}

As indicated above, for the two roots between $\Omega=\pm1$ being
complex the value of $B$ in eq.(\ref{eq:DispEq3}) must be below
its critical value:

\begin{equation}
B<B_{0}\equiv\frac{1}{4}\left(1+K^{1/3}\right)^{3}\label{eq:B0}
\end{equation}
The strategy behind finding these two complex roots is very simple.
First, we eliminate from consideration the remaining, allways real
roots that satisfy the condition $\left|\Re\Omega\right|>1$, by writing 

\[
\Omega=p+iq
\]
and requiring $q\neq0.$ This is a crucial requirement without which
the relation between $p$ and $q$ in eq.(\ref{eq:qSq}) below would
be meaningless. By separating the imaginary part of the l.h.s. of
eq.(\ref{eq:DispEq3}) we obtain this relation as follows:

\begin{equation}
q^{2}=\sqrt{1-p^{2}}\frac{\left(1-p\right)^{3/2}-\sqrt{K}\left(1+p\right)^{3/2}}{\sqrt{K}\sqrt{1-p}-\sqrt{1+p}}\label{eq:qSq}
\end{equation}
Now, we need to derive an equation for $p$, by using the real part
of eq.(\ref{eq:DispEq3}). After some straightforward algebra and
elimination of $q$ with the help of the above relation, we obtain

\begin{equation}
\sqrt{1-p^{2}}=\frac{2\sqrt{K}}{K+1-4Bp^{2}}\label{eq:Forp}
\end{equation}
As may be seen, this is a cubic equation for $p^{2}$ which is considerably
easier to solve than the original eq.(\ref{eq:DispEq3}). Let us make
the following transformation from $p$ to $\zeta$:

\begin{equation}
p^{2}=a\left(\zeta+1\right)+1,\;\;\;{\rm where}\;\;\;\;a=\frac{K+1}{6B}-\frac{2}{3}\label{eq:pSqRep}
\end{equation}
The equation for $\zeta$ then reads

\begin{equation}
3\zeta-4\zeta^{3}=1+\frac{K}{B^{2}a^{3}}\equiv Q\label{eq:zeta}
\end{equation}
Depending on the value of its r.h.s., $Q$, the last equation can
be solved by one of the following three substitutions:

\begin{equation}
\zeta=\begin{cases}
\sin\vartheta, & \left|Q\right|\leq1\\
\mp\cosh\phi,\;\;\;\;\; & Q\gtrless\pm1
\end{cases}\label{zetaCases}
\end{equation}
In particular, for $\left|Q\right|\leq1$ one obtains 

\begin{equation}
p^{2}=1+\left(\frac{K+1}{6B}-\frac{2}{3}\right)\left\{ \sin\left[\frac{1}{3}\sin^{-1}\left(1+\frac{6^{3}KB}{\left(K+1-4B\right)^{3}}\right)+\frac{2\pi}{3}\right]+1\right\} \label{eq:p2sin}
\end{equation}
Actually, the domain of function $Q\left(B\right)$ where $\text{\ensuremath{\left|Q\right|\le}1}$
does not need to be addressed any further. Indeed, in this region
either $B<0$ or $B>B_{0}$, Fig.\ref{fig:QofB}. The former case
is clearly impossible, eq.(\ref{eq:Bdef}). In the case $B>B_{0}$,
as we know, all four roots of eq.(\ref{eq:DispEq3}) are real and,
therefore, irrelevant to our analysis. 

The two remaining possibilities with $\left|Q\right|>1$ in eq.(\ref{zetaCases})
can be unified under one formula

\begin{equation}
p^{2}=\frac{1}{3}+\frac{K+1}{6B}-\left|\frac{K+1-4B}{6B}\right|\cosh\left[\frac{1}{3}\cosh^{-1}\left|1+\frac{6^{3}KB}{\left(K+1-4B\right)^{3}}\right|\right]\label{eq:p2ch}
\end{equation}
which provides the real part of the frequencies of both damped and
unstable modes. The imaginary parts (including the growth rate of
the unstable mode) can be obtained from eq.(\ref{eq:qSq}). Observe
that $q^{2}\left(-p,1/K\right)=q^{2}\left(p,K\right)$. In fact, there
are, only two roots with $q\neq0$, not four, as one might infer from
eqs.(\ref{eq:p2ch}) and (\ref{eq:qSq}) with no recourse to the properties
of the expression for $q^{2}$. This observation suggests the following
choice of the sign of $p$ when taking a square root from the expression
on the r.h.s. of eq.(\ref{eq:p2ch}). For $K<1$ we take the root
$p<0$, as to make $q^{2}$ positive. For $K>1$ we then choose $p>0.$
After the correct sign of $p$ is chosen, there are two complex conjugate
roots with the imaginary parts given by eq.(\ref{eq:qSq}). The two
remaining roots are always on the real $\Omega$- axis, since $B,K>0$.
These roots can easily be written in explicit form by factoring out
the two complex roots from the full quatric equation, eq.(\ref{eq:DispEq3}).

Eqs.(\ref{eq:p2ch}) and (\ref{eq:qSq}) thus provide the real and
imaginary parts of the unstable mode in terms of $K$ and $B$, eqs.(\ref{eq:Kdef},\ref{eq:Bdef}).
While $K$ is straightforwardly related to the wave propagation angle
with respect to the magnetic field, the parameter $B$ depends on
the main component of the wave vector, $k_{y}\gg k_{z}$ in a more
complicated fashion. We will thus use the dimensionless wave number
$b=\left|k_{y}\right|V_{Te}/\omega_{ce}$ instead of $B$ in presenting
the results for the frequency of the wave and for its growth rate
in eqs.(\ref{eq:p2ch}) and (\ref{eq:qSq}), respectively. For convenience,
we rewrite $B\left(b\right)$ using the following two parameters

\[
\Delta=\frac{1}{4}\frac{T_{i}}{T_{e}}\left(\Omega_{e}-\Omega_{i}\right)^{2},\;\;\;\nu=\frac{\omega_{ce}^{2}}{\omega_{pe}^{2}}
\]
So,

\[
B\left(b\right)=b\left[\nu b+1-I_{0}\left(b\right)e^{-b}\right]\Delta
\]
Furthermore, the normalization of $\Omega$ contains $k_{y}$ which
was necessary to minimize the number of parameters in mathematical
treatment of the dispersion eq.(\ref{eq:DispEq3}). Now we can return
to the physical frequency $\omega$ and growth rate $\Im\omega$,
and express the results in terms of $p$ and $q$ as follows

\begin{equation}
\Re\omega=\frac{k_{y}}{2}\left[\left(V_{Ey}-V_{di}\right)p+V_{Ey}+V_{di}\right]\label{eq:Reomega}
\end{equation}

\begin{equation}
\Im\omega=\frac{k_{y}}{2}\left(V_{Ey}-V_{di}\right)q\label{eq:Imomega}
\end{equation}
Here, the dimensionless frequency $p$ and growth rate $q$ are still
given by eqs.(\ref{eq:p2ch}) and (\ref{eq:qSq}). It is interesting
to note that the real part of the frequency contains a convective
contribution of combined drift speeds of electrons and ions, $V_{di}+V_{Ey}$.
The second contribution is made by the relative ion-electron motion
$V_{Ey}-V_{di}$ which is solely responsible for the growth rate. 

Shown in Figs.\ref{fig:pANDq1} and \ref{fig:pANDq2} are the components
of the frequency, $bp$, and growth rate, $bq,$ as functions of $K$
and $b$. These results characterize the dispersive and stability
properties of LH waves and are key ingredients of the anomalous transport
coefficients. The constant part of the phase velocity, $V_{ey}+V_{di}$
is subtracted from the surface plots. Now we can verify if the obtained
solutions are consistent with simplifications made in deriving eq.(\ref{eq:DispEq2})
and how they are related to experiment conditions.

\subsection{Verification and Discussion of the Solutions}

While eq.(\ref{eq:DispEq2}) has a generic form in which it is broadly
applied to many systems with counterstreaming plasmas and beams in
plasmas, there are limitations, both general and specific to the case
of colliding plasmas we consider here. First, eq.(\ref{eq:DispEq2})
describes local wave generation, not strictly applicable to inhomogeneous
and finite plasmas. Indeed the colliding plasmas are interacting and
mixing in a relatively thin layer, are inhomogeneous, and, in addition,
can be subject to macroscopic instabilities, such as rippling of the
plasma mixing layer \cite{Dieckmann2018PPCF...60a4014D}. Nevertheless,
because of a very short scale of the LHDI instability ($\sim\rho_{e}$),
the requirement $kL_{s}\gg1$ is satisfied and the local approximation
can be adopted ($L_{s}$ is the width of the collision layer, sec.\ref{subsec:Analysis-of-Colliding}).
With regard to the rippling effects, it indeed seems to be observed
in some experiments within a small part of the shocked layer, though
\cite{Sarkisov2005}. The field reconnection effects may play some
role in this as the rippling is more pronounced on the anode side
of the layer, where the electron current must be stronger, as they
are accelerated by the reconnection electric field in $y-$ direction.
On the other hand, recent experiments specifically designed to study
the reconnection layer in colliding flows do not show noticeable rippling
of the reconnection layer \cite{Suttle2018PhPl...25d2108S}.

Further constraints imposed on eq.(\ref{eq:DispEq2}) stem from the
hydrodynamic treatments of ions. So, their velocity dispersion should
be limited, even though their motion is randomized upon entering the
shocked plasma layer. Therefore, we have to impose the condition $\omega\gg kV_{Ti}$
which, in essence, means $V_{Ey}\gg V_{Ti}.$ We will consider the
electric field structure in the shocked plasma layer in detail further
in the paper, but for the purpose of specifying the above restriction
we can estimate $V_{Ey}=cE_{x}/B\sim c\phi_{s}/L_{s}B$, where $\phi_{s}$
is the electrostatic potential built up in the shocked plasma layer.
It can be estimated from the flow deceleration requirement, $e\phi_{s}\sim m_{i}u^{2},$
where $u$ is the flow speed upstream from the shocked layer. Combining
both inequalities, we obtain the following simple condition

\begin{equation}
\frac{\rho_{e}}{L_{s}}\ll1\ll\frac{\rho_{i}}{L_{s}}\label{eq:ApplCondiDispEq}
\end{equation}
where $\rho_{i}$ is the ion Larmor radius and we have substituted
the randomized ion velocity, $V_{Ti}\sim u$. Potentially problematic
may be the right part of this condition. However, the experiment \cite{Sarkisov2005}
indicates that the corona outflow speed $u\sim100$ km/s, while the
magnetic field remains in the range of $10$ kG. These numbers yield
for the ion Larmor radius $\rho_{i}\sim u/\omega_{ci}\sim10^{-1}A/Z$
cm, where $A$ and $Z$ are the mass and charge numbers of the ion.
Since the shocked layer in the experiment is shown to be significantly
narrower than the gap between the wires, $a=2$ mm, the condition
in eq.(\ref{eq:ApplCondiDispEq}) holds up, at least for incompletely
ionized ions with $A\gg1$. The recent reconnection dedicated experiments
\cite{Suttle2018PhPl...25d2108S} reach significantly higher magnetic
fields but also much stronger ion heating is observed, so the above
condition appears to be met as well. We also note here, that in the
interest of light notation, we do not include mass and charge numbers
$A,Z$ of the ions in the dispersion equation. They can be easily
restored by replacing $m_{i}\to Am_{i}$ and $e\to Ze$ in terms,
corresponding to the ion contribution.

\begin{figure}
\includegraphics[scale=0.6]{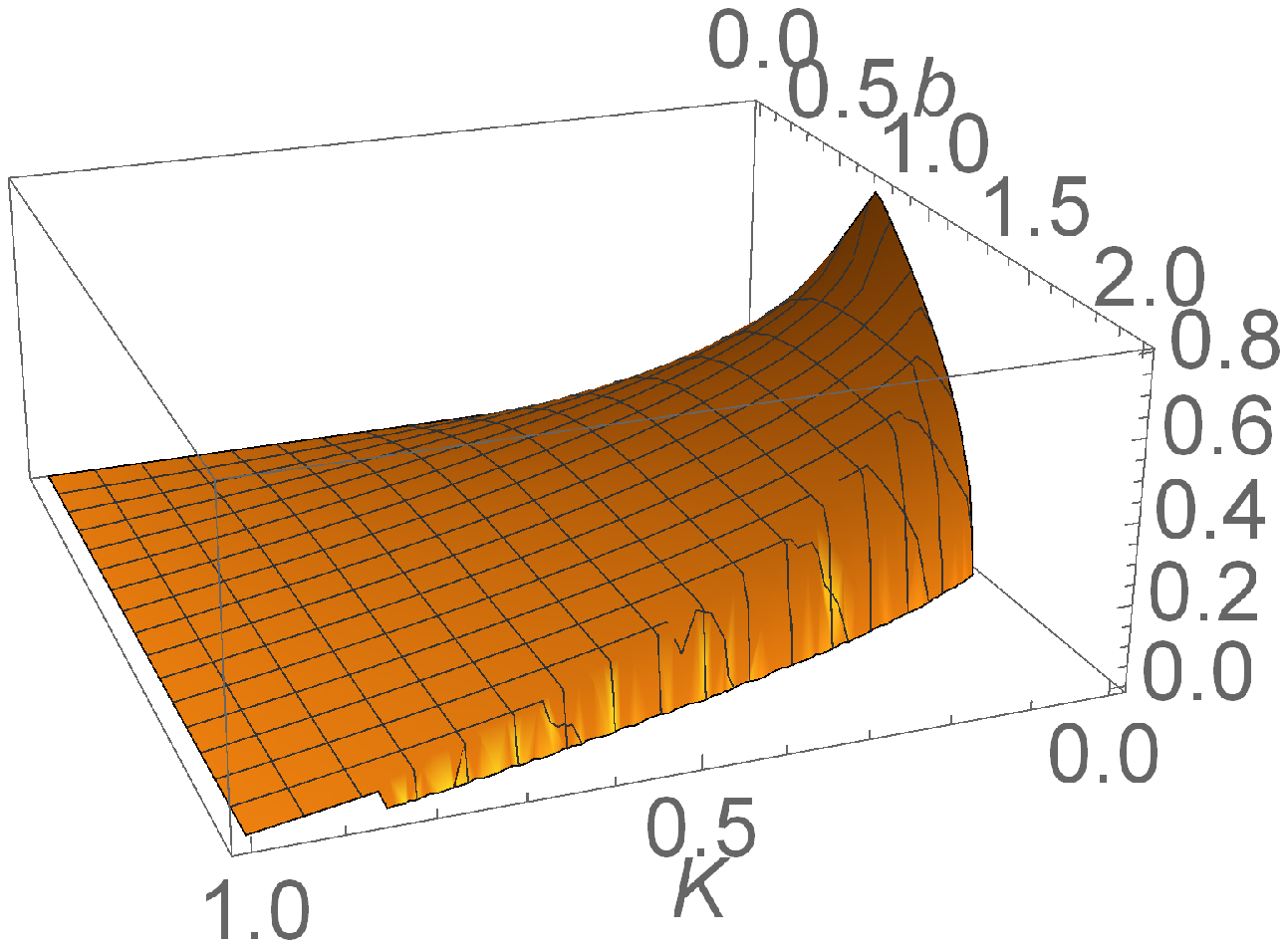}\includegraphics[scale=0.6]{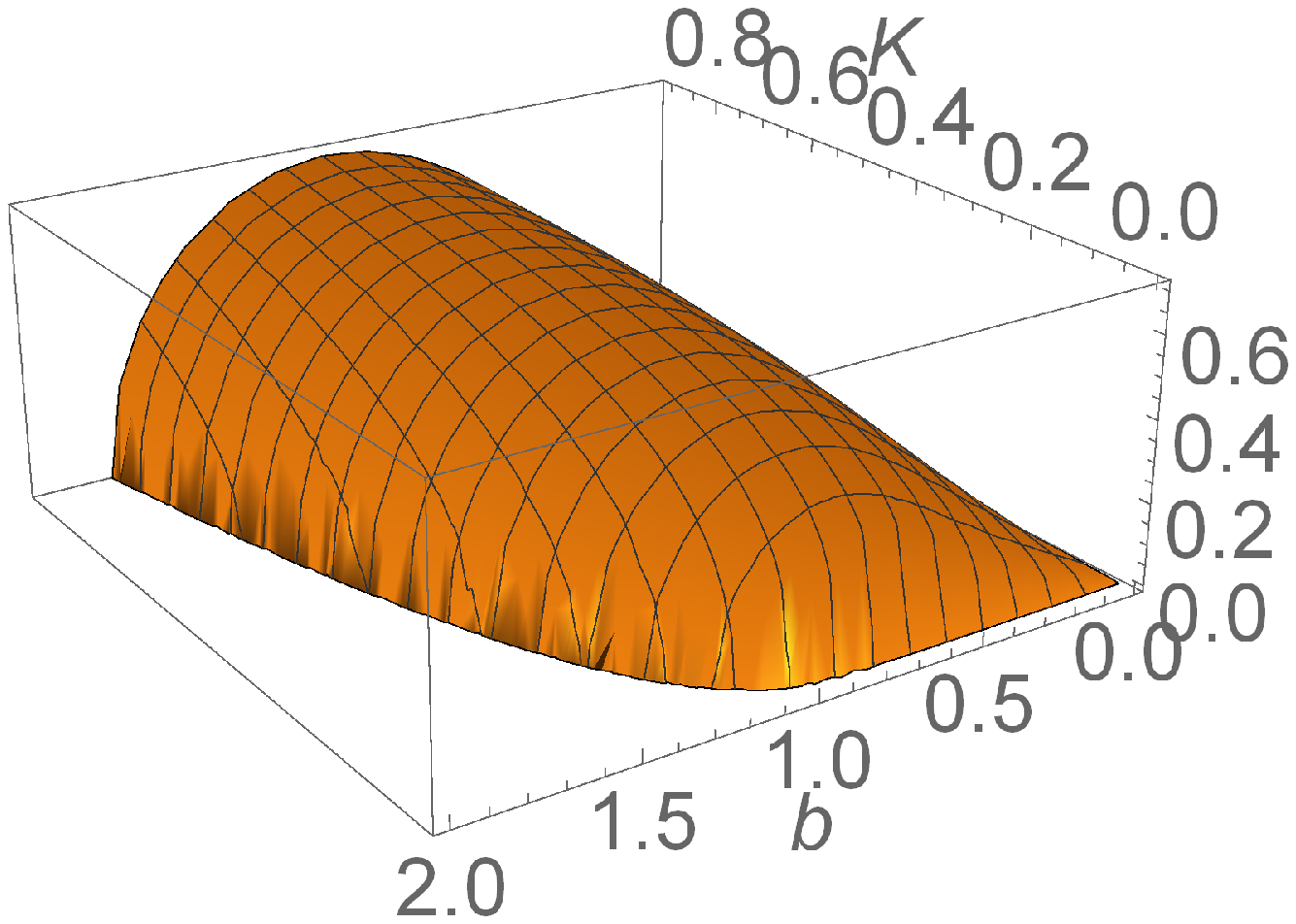}

\caption{Left panel: $bp$, a dimensionless dependence of the real part of
frequency $\omega$, as a function of $b$ and K. Shown is the unstable
region on the plane $\left(b,K\right)$, where $\Im\omega>0$, eq.(\ref{eq:Reomega}),
for $\Delta=1$ and $\nu=0.2$.\protect \\
Right panel: $bq$, the imaginary part of $\omega$, using the same
normalization as $bp$.\label{fig:pANDq1}}
\end{figure}

\begin{figure}
\includegraphics[scale=0.6]{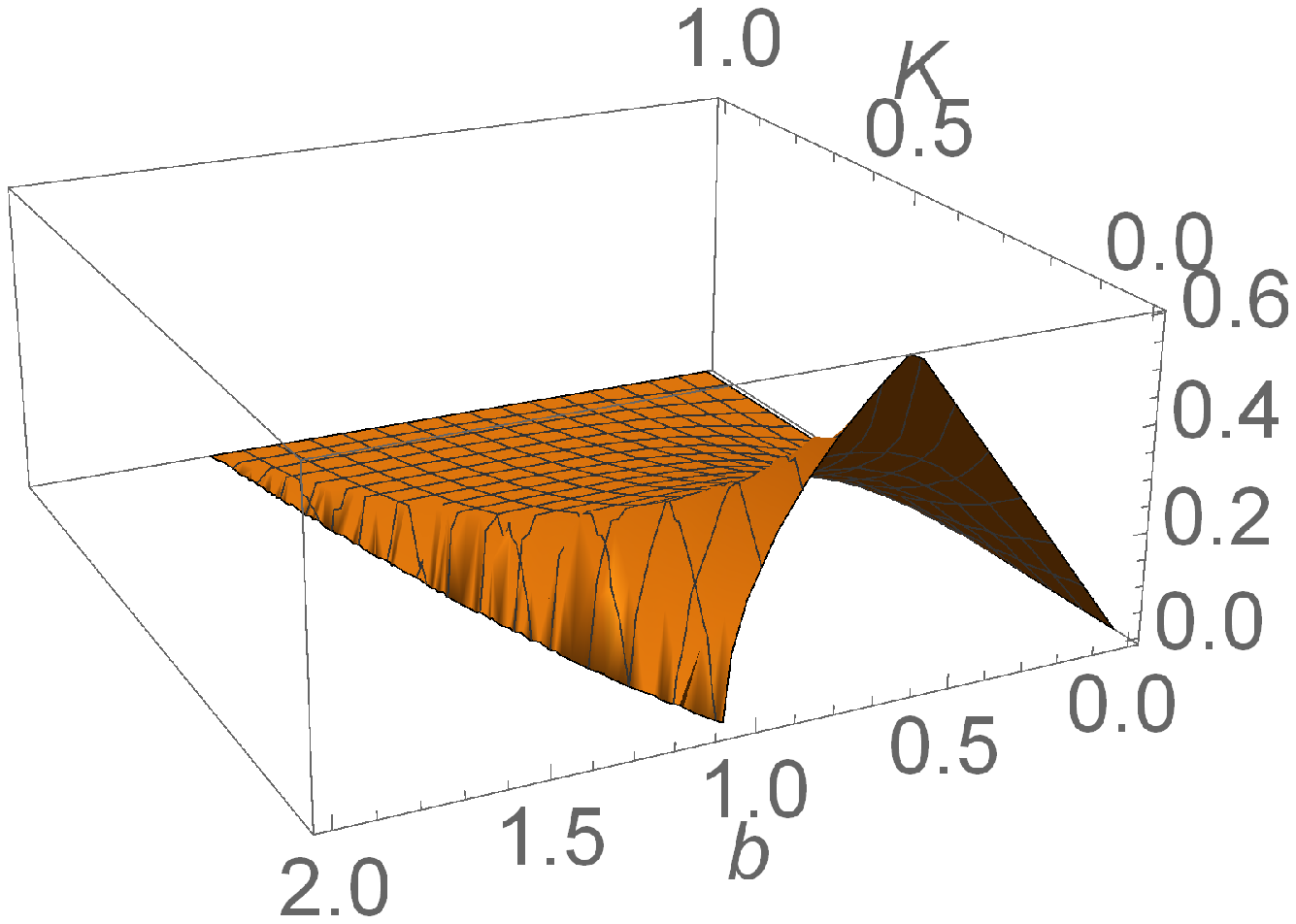}\includegraphics[scale=0.6]{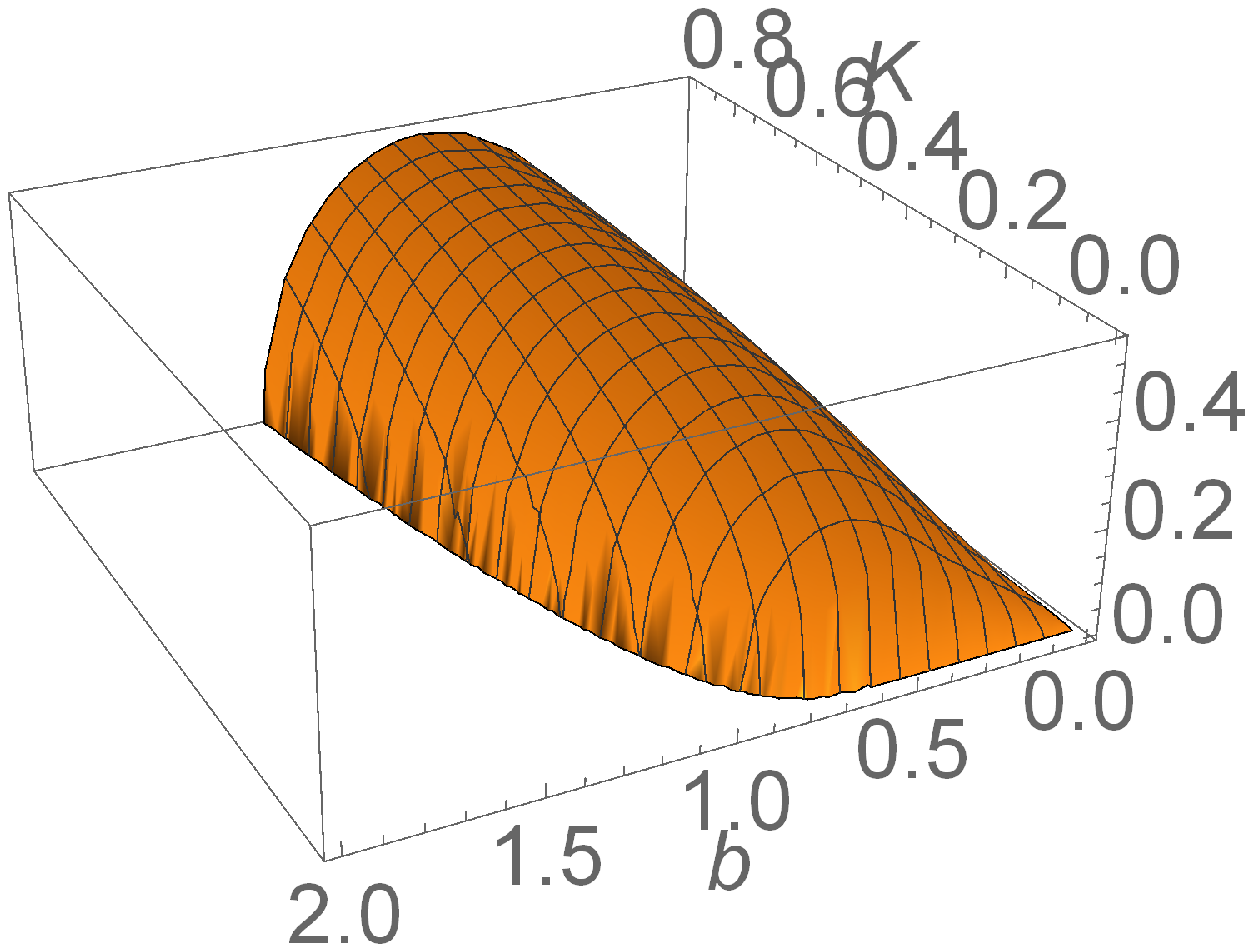}

\caption{The same as Fig. but for $\Delta=2.$\label{fig:pANDq2}}
\end{figure}

To conclude our discussion of the applicability of the results, we
observe that the waves propagate in $y-$ direction, which is favorable
for the homogeneous plasma approximation. Besides, the phase velocity
has a constant ($k$- independent) component $V_{Ey}+V_{di}$ that
remains finite even if the variable component $\sim p$ becomes small
for $K\sim1$. Note that the signs of the both drift velocities $V_{Ey}$
and $V_{di}$ coincide as both the magnetic fields around each wire
and the directional ion motion are associated with parallel currents.
This makes the condition in eq.(\ref{eq:ApplCondiDispEq}) easier
to fulfil. Moreover, even when the real part of the frequency, associated
with p-dependent contribution is small, the imaginary part of $\omega$
becomes large at $K\sim1$, thus reinforcing the condition in eq.(\ref{eq:ApplCondiDispEq}).
Note that the most unstable waves are those with $K\sim1$ (almost
perpendicular to $B$), $k_{z}/k_{y}\sim\sqrt{m/M}$. As $p<1$ (Figs.\ref{fig:pANDq1}-\ref{fig:pANDq2}),
they propagate in the electron and ion drift direction (positive $y$-
direction), according to eq.(\ref{eq:Reomega}). 

\section{Macroscopic Properties of Colliding Flows\label{sec:Macroscopic-Properties-of}}

In the preceding section, we obtained an exact formula for the growth
rate of LH-drift instability occurring in the region where two concentric
outflows from two exploding wires collide. They stream sideways from
the collision \foreignlanguage{american}{center} line between the
wires at $x=z=0$. The purpose of this section is to characterize
the resulting flow pattern whose parameters enter the stability analysis
of the preceding section. 

\subsection{The Overall Flow Configuration}

The analysis in the previous section indicates that the electron $\boldsymbol{E\times B}$
-drift in the flow collision region powers a strong LHDI instability.
The drift is supported by the electrostatic field $E_{x}$ built up
in the flow around the symmetry plane (magnetic zero). The $E\times B$
drift is then in $y$- direction, $V_{Ey}=cE_{x}/B$. It is more important
than electron and ion drifts associated with the external and reconnection
field $E_{y}$ for the following two reasons. First, the latter is
narrowly concentrated in the magnetic null plane. This region is only
$\sqrt{\rho_{e}L}$- wide, where $L=B/\left(dB/dx\right)$ is the
neutral layer characteristic scale which is obviously the same as
the flow collision region. By contrast, the $E_{x}$ field is supported
by the flow collision and associated charge separation caused by the
inertial forces in the colliding flows. Second, the experimental information
obtained from the explosive evaporation of specifically two wires
\cite{Sarkisov2005} point to a rapid voltage collapse, thus indicating
a short-circuiting of $E_{y}$ field. 

To get a grasp of the flow, it is useful to temporarily replace the
two colliding flows, Fig.\ref{fig:Colliding-flows-on}, by one, that
flows out concentrically from the point $x=-a$ and collides with
a reflecting wall placed at the symmetry plane at $x=0.$ A necessary
disclaimer to make here is that an important and likely observable
\cite{Sarkisov2005} rippling of the magnetic null surface is not
captured by this simplified treatment. Apart from this limitation,
and assuming a steady state with the conventional symmetry arguments,
the two flow configurations are equivalent. If, in addition, the plasma
inflow was strictly parallel to the $x-$ axis rather than concentric,
after having hit the wall, the flow could be described as a typical
numerical setup to simulate the shock waves. Namely, upon a specular
reflection off the wall, the reflected flow couples to the incident
one and a shock wave forms\footnote{Note that in the case of the equivalent colliding flow configuration,
the role of reflected ions is taken by the ions from the opposite
flow.}. Its structure and speed at which it propagates away from the wall
will depend on the Mach number of the inflowing plasma. This parameter
may be assumed to be larger than unity (so, the shock must form) but
not strongly so, as the adiabatic cooling of the cylindrical flow
scales as $T\propto r^{1-\gamma}$, where $\gamma$ is the adiabatic
index of the coronal phase. The cooling starts from about $r\sim10^{-3}-10^{-2}$cm
and ends at $r\sim a=10^{-1}$cm, e.g., in the experiment \cite{Sarkisov2005}.
Although the flow is magnetized and the shock will be of a magnetosonic
type, the plasma ram pressure typically exceeds the magnetic pressure
by a significant factor ($\sim10$) \cite{Sarkisov2005}, so we can
consider an electrostatically, rather than magnetically, dominated
ion-acoustic shock instead. 

Regardless the shock type, its important aspect is a critical Mach
number (at which it starts reflecting the upstream ions). Its exact
value for the ion-acoustic shocks significantly depends on the electron
distribution. For Boltzmannian electrons, $M_{cr}\approx1.6$ \cite{Sagdeev66},
while for adiabatically trapped electrons $M_{cr}\approx3.1$, \cite{Gurevich68}.
The choice between these two models should be made depending on the
macroscopic consequences of the LH instability. It is reasonable to
expect that this instability will result in a strong electron heating
and spatial diffusion, in particular. 

From the above consideration the following picture emerges: two coronal
flows, directed radially from the points $x=\pm a$, collide at $x=0$
and couple (presumably by a two-stream, Buneman instability). Then,
they slow down and decline sideways along the $z$- axis from the
flow stagnation line, $x=z=0$. Note, that although we do not consider
this initial two-stream instability here, it can be described by the
same type of dispersion equation as eq.(\ref{eq:DispEq3}), studied
in detail in Sec.\ref{subsec:Instability-Analysis}. The forces that
slow down and deflect the flow originate from an electrostatic potential
$\phi$ built up in response to the interaction of colliding flows
and a pondermotive potential, $\Phi$. It is connected with the unstable
oscillations, discussed in Sec.\ref{subsec:Instability-Analysis}.
In plasmas at equilibrium, the two potentials are usually related:
$\phi=-T_{i}\Phi/\left(T_{i}+T_{e}\right)$. In the case under consideration,
the ion inertial forces violate this simple balance. Evidently, the
flow must remain supersonic far away from the stagnation point, since
at large distances it becomes equivalent to a single wire outflow.
So, the downstream subsonic flow must transit through the sonic point
again while flowing out from the plane of the wires (in $z-$ direction).
Furthermore, two shocks are formed, as we show below, roughly parallel
to the mid-plane and propagate into the respective flows, Fig.\ref{fig:Colliding-flows-on}.
Technically, we can look for a steady state shock solution as long
as the inflowing plasma can be considered stationary. We assume that
this is the case and do not consider here the arrival of the wire
core material that is expected at later times \cite{Sarkisov2005}.

\subsection{Simplified Equation for the Flow}

\begin{wrapfigure}{o}{0.5\columnwidth}%
\includegraphics[bb=0bp 350bp 612bp 792bp,scale=0.47]{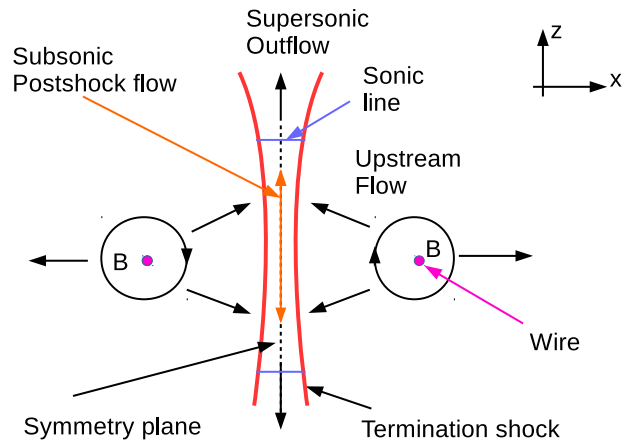}

\caption{Colliding flows on $x-z$ plane.\label{fig:Colliding-flows-on}}
 \end{wrapfigure}%

We will write the equations for colliding flows, using the following
normalized variables. The flow velocity will be measured in the units
of ion-sound velocity, $C_{s}=\sqrt{T_{e}/M}$, the potentials (both
electrostatic and pondermotive, $\phi$ and $\Phi$) in units of $T_{e}/e$.
The density is normalized to that of the corona, $n_{0}$, near the
surface where it starts to rise due to the flow collision, but where
the both potentials are still at zero values. This region is just
upstream of the shocks, standing in the flows between the wires. Since
this density changes along the $z-$ coordinate we take its value
at $z=0$ for the normalization purpose. Writing the Poisson equation
below, we measure the lengths in the units of Debye length, $\lambda_{e}=\sqrt{T_{e}/4\pi e^{2}n_{0}}$,
while other equations are scale-invariant. For a stationary flow,
they can be written in the following way

\begin{equation}
\frac{1}{2}\left(v_{x}^{2}+v_{z}^{2}\right)+\phi+\frac{T_{i}}{T_{e}}\ln n_{i}=\frac{1}{2}w_{0}^{2}=const\label{eq:Bern}
\end{equation}

\[
\frac{\partial v_{x}}{\partial z}-\frac{\partial v_{z}}{\partial x}=0
\]

\begin{equation}
\phi+\Phi-\ln n_{e}=0\label{eq:adiabEl}
\end{equation}

\begin{equation}
\frac{\partial}{\partial x}n_{i}v_{x}+\frac{\partial}{\partial z}n_{i}v_{z}=0\label{eq:Contin}
\end{equation}

\begin{equation}
\Delta\phi=e^{\phi+\Phi}-n_{i}\label{eq:Poiss}
\end{equation}
Several remarks about the above equations are in order here. First,
we have replaced the equation of motion for the ions by the Bernoulli's
integral in the first equation above, assuming that the flow remains
irrotational (second equation) also after it passes the shock. An
alternative, but in that regard equivalent, possibility is that for
a relatively low-Mach number flow the transition from the upstream
to downstream region is smooth. In this case the last equation can
be replaced by a simple quasi-neutrality condition, $n_{e}=n_{i}$.
Given the quasi-homogeneous flow upstream of both shocks, that are
relatively flat as we argue below, the curl-free condition, $\partial v_{x}/\partial z-\partial v_{z}/\partial x=0$,
is a plausible approximation \cite{Kevlahan97}. It is essential for
closing the system of the three equations that follow. The pondermotive
potential, $\Phi$, needs to be determined separately. We will calculate
it in a future study based on the growth rate obtained in Sec.\ref{subsec:Instability-Analysis}
and on the overall flow structure addressed in this section. 

\begin{wrapfigure}{o}{0.5\columnwidth}%
\includegraphics[clip,scale=0.3]{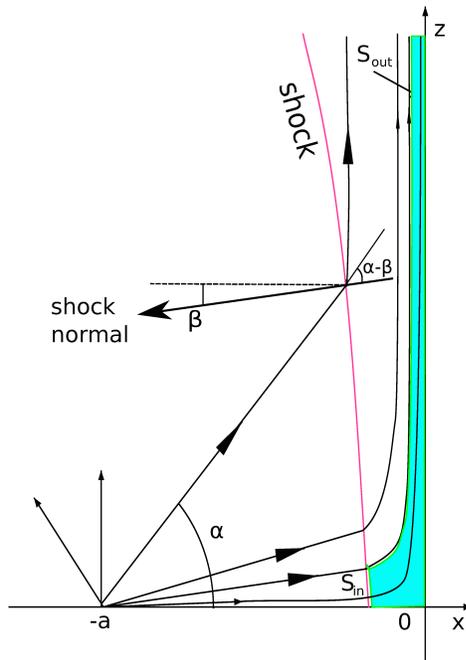}

\caption{Plasma flow past the right angle at $x=z=0$, that is equivalent to
the colliding outflows under two-way symmetry: $x\to-x,$ $z\to-z$.\label{fig:Plasma-flow-past}}
\end{wrapfigure}%

As the quasineutrality condition is not yet implemented in the Poisson
equation, the above equations capture the electrostatic shock transition.
However, the ion density $n_{i}$, except for the vicinity of the
shock transitions, is indeed very close to the electron density, $n_{e}$
(quasineutrality condition), as already mentioned. The equation of
motion for electrons is replaced by a simple balance condition between
the pondermotive, electrostatic and pressure forces. The balance results
in a Boltzmannian distribution in the total potential $\phi+\Phi$,
as may be seen from eq.(\ref{eq:adiabEl}). An alternative model for
the electron distribution near the shock front assumes their adiabatic
trapping in the shock potential \cite{Gurevich68}. In this case the
critical Mach number beyond which the ions begin to reflect off the
shock is about a factor two higher than in the case of Boltzmannian
electron distribution, where $M_{cr}\approx1.6$. These values have,
however, been obtained for one-dimensional ion-acoustic shocks and
can be used only as guide lines in the two-dimensional flow analysis. 

To minimize the number of parameters in the above system of equations
(only two of them remain, $T_{e}/T_{i}$ and $w_{0}$), we have chosen
the Debye length, $\lambda_{D}$, for the space scale. This is a natural
scale to describe the shock transition and it is too short for the
rest of the flow. However, as we apply the quasi-neutrality condition
(outside of the shock transition we can set $\Delta\phi=0$) the above
system becomes scale-invariant, and we can use any other convenient
scale, e.g., the distance between the wires, $2a$, etc. To conclude
this introduction of the equations used in this section we note that
some of the auxiliary notations from Sec.\ref{sec:A-Model-for} will
be reused here, which should not cause confusion.

The 2D-flow described by the above equations is depicted in Fig.\ref{fig:Colliding-flows-on}.
The back-transitions to the supersonic flows, shown by the two ``sonic
lines,'' logically follows from the equivalence of the double- and
single-wire flows at large distances from the origin ($r\gg a$),
as we pointed out earlier. An equivalent single-wire coronal flow
is presumed to be marginally supersonic, according to the corona parameters
inferred from the experiments \cite{Sarkisov2005}. 

\subsection{Analysis of Colliding Flows\label{subsec:Analysis-of-Colliding}}

Referring to Fig.\ref{fig:Colliding-flows-on} again, let us focus
on the region enclosed by the two shocks and two sonic lines. We need
to calculate the width $2X_{s}\left(x\right)$ of the post shock region
between the shocks at $\text{\ensuremath{\left|x\right|}<}X_{s}\left(z\right)$,
as it determines the scale of the macroscopic electric field $E_{x}$
that, in turn, sets the growth rate of the LHDI. This quantity needs
to be related to the flow Mach number $w_{0}=\sqrt{v_{x}^{2}+v_{z}^{2}}$
(the velocities are already measured in units of $C_{s}$) and the
jump of electrostatic potential, $\phi_{s}\left(z\right)$, at the
termination shocks. This jump is similar to the maximum potential
of the conventional ion-acoustic soliton with a trailing oscillatory
structure \cite{Sagdeev66,MoisSagd63}. There are several important
distinctions, though.

First, we are dealing with a two-dimensional shock with gradually
changing flow angle from the normal incidence at $z=0$ to the increasingly
oblique shock geometry at $z\gtrsim a$. Besides, the shock weakens
with $z$ to merge into the radial supersonic outflow at large $z\gg a$,
beyond the outflow sonic line. Since the shock is generally oblique,
in applying the shock jump conditions we will allow only the normal
velocity component to jump while keeping the tangential one continuous
across the shock surface. Second, as the upstream flow cross the shock
at a variable distance from the centers of the outflows located at
$x=\pm a$, the upstream shock conditions also change. Finally, the
shock electrostatic potential, while possibly oscillating in a narrow
region (a few $\lambda_{De}$) behind the shock transition (as in
the ion-acoustic shocks with the ion reflection, studied in \cite{MSetal_IAshocks2016}),
on average it must gradually increase towards the mid-plane. This
is a symmetry requirement, since ions must progressively deviate from
the \foreignlanguage{american}{center-line} of the flow downstream
($z=0$). For the same reason, the flow must stagnate at the point
$x=z=0$. We will discuss this aspect of the shock in more detail
in Sec.\ref{subsec:Shock-geometry-and}.

Assuming the quasineutrality condition in the shocked part of the
flow, from eqs.(\ref{eq:Bern}-\ref{eq:Poiss}) we obtain the following
relation between the maximum electrostatic potential, $\phi_{{\rm max}}$,
and the Mach number, $w_{0}$:

\begin{equation}
\phi_{{\rm max}}=\frac{\frac{w_{0}^{2}}{2}-\frac{T_{i}}{T_{e}}\Phi_{0}}{1+\frac{T_{i}}{T_{e}}}\label{eq:fiMax}
\end{equation}
Here $\phi_{{\rm max}}$ and $\Phi_{0}$ relate to the point $x=z=0$.
It is seen that the micro-scale flow instability, studied in Sec.\ref{subsec:Instability-Analysis}
and entering the flow potential through the pondermotive potential
$\Phi$, may play a key role in mediating the electrostatic shock
and shaping the flow. This effect is expected to be more pronounced
in strong shocks where the ion velocity randomization upon shock crossing
(higher $T_{i}$) strongly dominates the electron shock thermalization.
At the same time, the electron heating should also result from the
lower-hybrid instability and will, on the contrary, diminish the role
of the pondermotive pressure. 

Our purpose here, however, is to understand the general characteristics
of the flow and estimate the shock stand-off distance, that is the
width of the shocked plasma layer, as a function of $z$. These characteristics
of the flow will provide information about the electrostatic potential,
that includes the shock jump condition and its post shock growth up
to its maximum at the flow stagnation point. The potential profile
is key in calculating the flow micro-instabilities, studied in Sec.\ref{sec:A-Model-for}.
It is clear that this potential slows down and compresses the colliding
flows as the internal and pondermotive pressures also do. However,
the maximum flow compression (downstream to upstream density ratio)
is limited by a factor depending on the polytrope index. 

\[
n_{2}/n_{1}=\left(\gamma+1\right)/\left(\gamma-1+2/w_{0}^{2}\right),
\]
where $w_{0}$ is the Mach number of the flow ahead of the shock.
Note, however, that due to radiation and ionization effects, especially
in dense plasmas, the polytrope index may be significantly lowered
compared to the familiar mono-atomic gas $\gamma=5/3$, e.g. down
to $\gamma\approx1.2$ \cite{Drake2018}. Indeed, according to the
observations \cite{Sarkisov2005}, the layer of compressed plasma
between the wires is rather thin. We therefore consider the electrostatic
potential to be the primary flow decelerator. As we will see below,
in this case the system given by eqs.(\ref{eq:Bern}-\ref{eq:Poiss})
is equivalent to the set of gasdynamic equations with the adiabatic
index $\gamma=1$.

That the electrostatic potential is capable of compressing a cold
ion flow ($T_{i}\ll T_{e}$) significantly stronger than the adiabatic
compression, can be illustrated by considering a simple ``water bag''
model. According to this model, the ions are distributed in the normal
to the shock speed $v$ in the following way $f_{i}$$\left(v\right)=\eta\left(v-v_{1}\right)\eta\left(v_{2}-v\right)$.
Here $\eta$ is the Heaviside unit function and $v_{2}>v_{1}$. By
virtue of Liouville's theorem and particle energy conservation, after
passing over a potential jump of the height $\phi$, the ion distribution
becomes $f_{id}$$\left(v\right)=\eta\left(v-\sqrt{v_{1}^{2}-2\phi}\right)\eta\left(\sqrt{v_{2}^{2}-2\phi}-v\right)$.
So, the flow compression ratio can be written as: 

\[
\frac{\sqrt{v_{2}^{2}-2\phi}-\sqrt{v_{1}^{2}-2\phi}}{v_{2}-v_{1}}
\]
The flow compression reaches its maximum when the slowest particles
with $v=v_{1}$ begin to reflect, that is for $\phi=v_{1}^{2}/2$,
we can write

\[
\sqrt{\frac{v_{2}+v_{1}}{v_{2}-v_{1}}}\sim\sqrt{V_{i}/V_{Ti}}
\]
where $V_{i}$ and $V_{Ti}$ denote the ion bulk and thermal velocities,
respectively. This compression may indeed considerably exceed that
of the conventional shock adiabat for a cold and fast ion flow thus
being more consistent with the observations \cite{Sarkisov2005}.
To further simplify the derivation we neglect the ion pressure contribution
in Bernoulli eq.(\ref{eq:Bern}), by assuming that the balance in
it is dominated by electrostatic and inertial forces on the flow.
Neglecting also the deviation from quasineutrality in the post shock
region, we rewrite the system given by eqs.(\ref{eq:Bern}-\ref{eq:Poiss})
as follows

\begin{equation}
v_{x}^{2}+v_{z}^{2}+2\phi=w_{0}^{2}=const\label{eq:Bern2}
\end{equation}

\[
\frac{\partial v_{z}}{\partial x}-\frac{\partial v_{x}}{\partial z}=0
\]

\begin{equation}
\frac{\partial}{\partial x}v_{x}e^{\phi}+\frac{\partial}{\partial z}v_{z}e^{\phi}=0\label{eq:Contin2}
\end{equation}
In view of the $n_{e}=n_{i}$ approximation made here, these equations
do not describe the shock transition and will only be applied to the
flow in the shocked plasma between the two termination shocks. In
the conventional gasdynamics, a termination shock forms where the
ram pressure of the outflow is equilibrated by the pressure of the
ambient medium. In the case of colliding flows considered here the
pressure of shocked plasma between the two shocks plays the role of
the pressure of the ambient medium. In our model, however, the electrostatic
potential appears to be a better mediator between the counter-streaming
flows. Most of its rise in the stagnating flow occurs immediately
at the shock transition. The cross shock potential can be obtained
from the local treatment in a one-dimensional shock approximation.
Now we focus on the two-dimensional \foreignlanguage{american}{post-shock}
flow.

Introducing the flow potential $\chi$ by 

\begin{equation}
v_{x}=\partial_{x}\chi,\;\;\;v_{z}=\partial_{z}\chi,\label{eq:chiDef}
\end{equation}
we rewrite eqs.(\ref{eq:Bern2}-\ref{eq:Contin2}) in the following
form

\begin{equation}
\left|\nabla\chi\right|^{2}+2\phi=w_{0}^{2}\label{eq:PsiFi1}
\end{equation}

\begin{equation}
\Delta\chi+\nabla\chi\cdot\nabla\phi=0,\label{eq:PsiFi2}
\end{equation}
where $\nabla=\left(\partial_{x},\partial_{z}\right).$ By the symmetry
of the flow, it is sufficient to consider the flow in any quadrant
of the $x-z$ plane and the adjacent part of the respective shock,
Fig.\ref{fig:Plasma-flow-past}. The flow is then past the interior
of the $90^{\circ}-$ angle, e.g. from the positive $x$ direction
at $x<0,$ $z\approx0$ turning to the positive $z-$ direction at
$z>0,\;x\approx0$. The boundary conditions at the walls of the corner
are dictated by the symmetry requirements at $x=0$: $\partial_{x}\chi=0$,
$\partial_{x}\phi=0,$ and similar conditions at $z=0:$ $\partial_{z}\chi=0$,
$\partial_{z}\phi=0$. Substituting $\phi$ from eq.(\ref{eq:PsiFi1})
into eq.(\ref{eq:PsiFi2}) one can obtain one equation for the flow
potential

\begin{equation}
\Delta\chi=\frac{1}{2}\nabla\chi\cdot\nabla\left|\nabla\chi\right|^{2}\label{eq:Psi3eq}
\end{equation}

The above system or, equivalently, the system given by eqs.(\ref{eq:Bern2}-\ref{eq:Contin2})
can be linearised by transforming it to a hodograph plane, e.g. \cite{LLFM},
on which dependent and independent pairs of variables interchange.
The problem is, however, in the boundary conditions at the shock whose
position is unknown beforehand and should be determined self-consistently
with the flow past the shock. This makes the application of hodograph
method rather problematic. We therefore begin with an approximate
analysis of the system given by eqs.(\ref{eq:PsiFi1}-\ref{eq:PsiFi2}). 

Referring to Fig.\ref{fig:Plasma-flow-past}, let us consider first
the flow near the corner $x=z=0$. As this is the stagnation point,
we can neglect nonlinear velocity terms, to the first order of approximation.
Specifically, the first term in eq.(\ref{eq:PsiFi2}) is of the order
of $v/L_{s}$ while the second term is $v^{2}\ll1$ times smaller.
Here $L_{s},$ stands for the distance between the shock and the flow
stagnation point, while the function $X_{s}\left(z\right)$, introduced
earlier, is related to $L_{s}$ by $L_{s}=X_{s}\left(0\right).$ The
flow potential near the stagnation point can then be approximated
by a harmonic function, $\Delta\chi=0$. Given the boundary conditions,
we can write the solution of the Laplace equation as

\begin{equation}
\chi\approx\frac{U_{0}}{2L_{s}}\left(z^{2}-x^{2}\right)\label{eq:PsiZ2minx2}
\end{equation}
Here the post shock velocity $U_{0}\equiv v_{x}\left(-L_{s},0\right)$
is taken at the mid-plane $z=0$. It will be specified later using
the shock jump conditions. The electrostatic potential near the stagnation
point can now be obtained immediately from eq.(\ref{eq:PsiFi1}):

\begin{equation}
\phi\approx\frac{w_{0}^{2}}{2}-\frac{U_{0}^{2}}{2L_{s}^{2}}\left(x^{2}+z^{2}\right)\label{eq:FiX2plusZ2}
\end{equation}
It follows that the flow changes its direction from predominantly
horizontal at $z<\left|x\right|$ to vertical at $z>\left|x\right|$,
deflected by the potential hump in eq.(\ref{eq:FiX2plusZ2}). 

To gain more insight into the shock geometry, let us integrate the
continuity eq.(\ref{eq:Contin2}) over a narrow region inside the
flow (filled region in Fig.\ref{fig:Plasma-flow-past}). This region
is confined by the two adjacent streamlines and the two lines transversal
to the flow that the flow crosses when entering and exiting the region.
Let the first transversal has a length $S_{in},$ being a part of
the shock surface. The second transversal, of the length $S_{out},$
is far away from the stagnation point, where the flow accelerates
to its preshock speed, that is beyond the sonic line. The contributions
to the integral from the two stream lines vanish, while the contributions
from the lines $S_{in}$ and $S_{out}$ remain. So, we obtain

\[
S_{in}e^{\phi_{s}}\sqrt{w_{0}^{2}-2\phi_{s}}=S_{out}w_{0}
\]
For shocks near criticality, where the cross shock potential $\phi_{s}$
is about to reflect the inflowing ions $w_{0}^{2}\approx2\phi_{s}$,
we arrive at a strong inequality $S_{in}\gg S_{out}$. We have taken
into account that the exponential factor is typically not very large.
For example, the critical Mach number of an ion-acoustic shock with
Boltzmannian electrons is $w_{cr}\approx1.6$, as we mentioned earlier.
For the adiabatically trapped electrons, though, this quantity is
higher, $w_{cr}\approx3.1$. The former model is, however, a better
choice for the case considered, as reasonably frequent electron-ion
collisions should maintain the local Maxwellian.

From the above relation we see that the flow tubes become asymptotically
narrower by a factor $\sim\left(1-2\phi_{s}/w_{0}^{2}\right)^{1/2}\ll1$
upon passing the corner at the stagnation point $x=z=0$. For an estimate,
we can assume the cross-section of the net inflow $S_{in}$ to be
$S_{in}\lesssim a$. Indeed, for larger $S_{in}$ the flow incidence
angle with respect to the shock normal becomes large enough for particle
reflection and these particles do not enter the post-shock space.
Thus, we obtain for the total width of the outflow $S_{out}\lesssim\left(1-2\phi_{s}/w_{0}^{2}\right)^{1/2}a\ll a$,
with an implication that the shock region width $L_{s}\ll a$, which
is consistent with the experiments \cite{Sarkisov2005}.

\subsection{Shock geometry and cross-shock potential\label{subsec:Shock-geometry-and}}

We have found the shocked plasma in colliding flows to form a thin
shell between two shocks terminating each flow. Now we wish to understand
the shape of this shell, which requires the knowledge of the flow
velocity between the shocks and the distribution of electrostatic
potential along the shock surface. Under the symmetry conditions and
constraints concerning the shocked plasma flow, that we discussed
and accepted earlier, we can replace this flow by an equivalent one.
This latter flow is between one shock and the two rigid walls making
the right angle, as depicted in Fig.\ref{fig:Plasma-flow-past}. 

The flow velocity on the downstream side of the shock surface can
be expressed through the shock jump conditions, Fig.\ref{fig:Plasma-flow-past}: 

\begin{equation}
v_{\tau}=w_{0}\sin\left(\alpha-\beta\right)\label{Vtau}
\end{equation}

\begin{equation}
v_{n}=\sqrt{w_{0}^{2}\cos^{2}\left(\alpha-\beta\right)-2\phi_{s}}\label{eq:Vn}
\end{equation}
They constitute the continuity of the tangential component of the
flow velocity and the jump of its normal component due to the cross-shock
potential $\phi_{s}$, under the conservation of particle energy during
the shock crossing. As in the previous subsection, $w_{0}$ denotes
the flow velocity ahead of the shock, while $v_{n}$ refers to the
normal velocity component downstream from the shock surface. The angles
$\alpha\left(z\right)$ and $\beta\left(z\right)$ refer to the flow
angle to the $x-$ axis ahead of the shock and the shock angle to
$z-$ axis, respectively. These angles are related to the distance
between the shocks separation,$2X_{s}\left(z\right)$, by

\[
\tan\alpha\left(z\right)=\frac{z}{a-X_{s}\left(z\right)}
\]

\[
\tan\beta\left(z\right)=\frac{dX_{s}}{dz}
\]
where $X_{s}\left(z\right)>0$ is the shock distance from the mid-plane
$x=0$, Fig.\ref{fig:Plasma-flow-past}. We have already discussed
the two remaining boundary conditions at the ``walls'' $x=0$ and
$z=0$. 

At the distances $z\gg L_{s}$ from the flow base line $z=0$, the
shocked flow is directed nearly vertically. This is dictated by the
alignment of the stream lines coming from the bottom parts of the
shock along the $z-$ axis. The sharp turn of the flow from $x$ aligned
to $z-$ aligned can be described by the flow structure and electrostatic
potentials already given in eqs.(\ref{eq:PsiZ2minx2}) and (\ref{eq:FiX2plusZ2}).
As we already argued, for shocks with sufficiently high Mach numbers,
when the ions are about to reflect, they move very slowly in the shock
normal direction. Therefore, the change of the flow direction will
occur in a narrow layer near the shock front. 

To begin with a preliminary sense of how the shock potential is distributed
along the shock surface, we make the following simplification. Instead
of resolving this narrow layer and matching the solution to the nearly
vertical flow between the shocks, we can collapse this region and
require that the flow becomes vertical immediately after the shock
crossing. That is, we impose the condition $v_{x}\left(-X\left(z\right),z\right)=0$
which translates into the following relation 

\begin{equation}
\cos\beta\sqrt{\cos^{2}\left(\alpha-\beta\right)-2\phi_{s}\left(z\right)/w_{0}^{2}}-\sin\left(\alpha-\beta\right)\sin\beta=0\label{eq:ShockGeom}
\end{equation}
The $z-$ component of the flow velocity can then be written as follows

\[
v_{z}=w_{0}\frac{\sin\left(\alpha-\beta\right)}{\cos\beta}
\]
One can also obtain a very simple expression for the $\phi_{s}\left(z\right)$
dependence where $z$ is sufficiently large, so that the flow incidence
angle is not small, $\alpha\gg\beta$. As we argued, the shock surface
must be well aligned with the vertical axis, so we may set $\beta\approx0.$
Under these conditions, from eq.(\ref{eq:ShockGeom}) we have

\begin{equation}
\phi_{s}\left(z\right)=\frac{1}{2}\frac{w_{0}^{2}}{1+z^{2}/a^{2}}\label{eq:fi-s-ofz}
\end{equation}
The electrostatic potential decreases with growing $z$ along the
shock front as to keep its strength close to criticality. The upstream
plasma impinging on the shock surface thus looses its horizontal velocity
component or even reflects from the shock at sufficiently large $z$,
when the incidence angle $\alpha$ is not too small. 

\subsection{Hodograph Equations\label{subsec:Hodograph-Equations}}

Now that we have a better grasp of the flow between the shocks, we
shall attempt at the analysis of its properties on a more accurate
basis. For this purpose, we shall map the nonlinear eqs.(\ref{eq:Bern2}-\ref{eq:Contin2}),
describing a plane-parallel flow on the $x,z$ -plane to the so-called
hodograph plane. In the new form, the equations will be linear. Meanwhile,
the original equations represent a special case of the gas of $\gamma=1$
adiabat (or isothermal flow, cf. eqs. {[}\ref{eq:Bern2}{]} and {[}\ref{eq:Contin2}{]}),
since the quasi-neutrality condition and the Boltzmann distribution
for electrons require the electrostatic potential to play the role
of the specific enthalpy of the gas in the following form: $\phi=\ln\rho$.
This does not make the hodograph transform derivation any different
from the conventional gas-dynamic case with $\gamma>1$. We therefore
provide only the key steps of this derivation in Appendix \ref{sec:Hodograph-Transform-App}.
As usual, the main problem in applying the hodograph method is in
the boundary conditions of the flow. 

\begin{figure}
\includegraphics[bb=20bp 570bp 612bp 792bp,clip,scale=0.9]{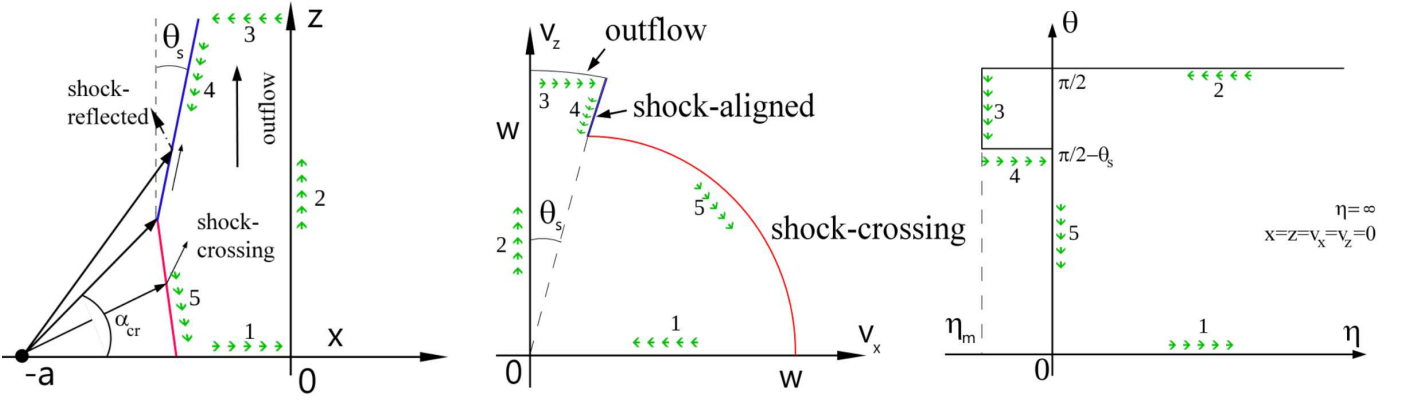}

\caption{Hodograph transform: physical flow plane $x,z$ (left panel); $v_{x},v_{z}$
plane (middle); $\eta,\vartheta$ plane (right panel). The solid lines
with arrows on the left panel schematically show the streamlines (see
also Fig.\ref{fig:Plasma-flow-past}). The green arrows indicate the
direction on the contour consisting of the segments 1-5 (see text).
\label{fig:Hodograph-transform:-physical} }
\end{figure}

We will study a specific part of the flow that is confined by the
shock surface and two axes of symmetry ($x=0,\;z>0$ and $z=0,\;x<0$),
as shown in Fig.\ref{fig:Plasma-flow-past}. The fourth line, that
closes the boundary of the flow under consideration, is drawn somewhat
arbitrarily in the outflow region where $z\gtrsim a$. The flow is
accelerated to transonic speeds in this region, $v_{z}\lesssim C_{s}$
($v_{z}\lesssim1$ in dimensionless variables that we use) while the
$v_{x}$ component becomes close to zero, as discussed in the preceding
subsection. We introduce this latter boundary in the flow because
we are not concerned here with the transition back to the supersonic
regime that occurs far away from the flow stagnation point. To proceed
on the hodograph derivation, in addition to the flow potential $\chi$
used in the previous subsection, we introduce a stream function $\psi$
in a usual way,

\begin{equation}
\rho v_{x}=\frac{\partial\psi}{\partial z},\;\;\;\rho v_{z}=-\frac{\partial\psi}{\partial x}\label{eq:PsiDef}
\end{equation}
Transforming from the $x,z$ dependent variables to $v_{x},v_{z}$,
we use polar coordinates on the flow velocity plane adhering to the
traditional hodograph notations $q,\vartheta$ for them:

\begin{equation}
v_{x}=q\cos\vartheta,\;\;\;v_{z}=q\sin\vartheta\label{eq:VelPolar}
\end{equation}
The hodograph equations take then the following (Chaplygin) form,
Appendix \ref{sec:Hodograph-Transform-App}: 

\begin{eqnarray}
\frac{\partial\chi}{\partial q} & = & \frac{q^{2}-1}{\rho q}\frac{\partial\psi}{\partial\vartheta}\label{eq:Chapl1}\\
\frac{\partial\psi}{\partial q} & = & \frac{\rho}{q}\frac{\partial\chi}{\partial\vartheta}\label{eq:Chapl2}
\end{eqnarray}
where, in accordance with eqs.(\ref{eq:Bern2},\ref{eq:Contin2})

\begin{equation}
\rho=\rho_{0}e^{-q^{2}/2},\;\;\;\rho_{0}=e^{w_{0}^{2}/2}\label{eq:rhoOfqDef}
\end{equation}
Again, we assume for simplicity that the plasma enters the shock with
constant density (we normalized it to unity) and constant supersonic
velocity $w_{0}>1$. In fact, this is not quite true and even though
the constant velocity assumption is plausible (as is approximately
the case in various supersonic outflows), the density is inversely
proportional with the distance from the source at $x=-a$, just to
conserve the flux, Fig.\ref{fig:Plasma-flow-past}. However, the most
important part of the flow is at relatively low values of $z\lesssim a$
since this is the region where the electric field drives the drift
in $y-$ direction and thus the lower-hybrid instability. Therefore,
we may ignore the foreshock variations in the flow in view of much
stronger changes in velocity and density occurring upon crossing the
shock. 

Formally, Chaplygin eqs.(\ref{eq:Chapl1}-\ref{eq:Chapl2}) are equivalent
to the original system given by eqs.(\ref{eq:Bern2}-\ref{eq:Contin2}).
Their two big advantages over the original system are that they are
linear and the variables $q,\vartheta$ are separable. As we mentioned
earlier, the price of these simplifications is in the boundary conditions,
since they must be formulated in terms of the flow velocity components
as new independent variables, rather than fixed geometrical boundaries.
At the same time, the specificity of our problem is that two of the
boundaries are not known regardless of the form of governing equations
in use. These are the shock surface function, $x=X_{s}\left(z\right)$
and the position of the sonic line. The latter, though, is of a lesser
concern for this study. 

These difficulties notwithstanding, our task can be simplified because
the large part of the flow is strongly decelerated upon passing through
the shock wave but not yet accelerated to the supersonic velocities
($q>1$), while proceeding in $z$ direction. Therefore, we can consider
the case of $q<1$ and observe some cancellations of the $q$-dependencies
in coefficients of eqs.(\ref{eq:Chapl1}-\ref{eq:Chapl2}) for $q\ll1$.
These cancellations allow us to evaluate the Chaplygin equations to
a system with constant coefficients, accurate to the order $q^{4}$.
Indeed, to this accuracy we can rewrite eqs.(\ref{eq:Chapl1}-\ref{eq:Chapl2})
as follows

\begin{eqnarray}
\frac{\partial\chi}{\partial q} & \approx & -\frac{1}{\rho_{0}q}e^{-q^{2}/2}\frac{\partial\psi}{\partial\vartheta}\label{eq:Chapl1-1}\\
\frac{\partial\psi}{\partial q} & = & \frac{\rho_{0}}{q}e^{-q^{2}/2}\frac{\partial\chi}{\partial\vartheta}\label{eq:Chapl2-1}
\end{eqnarray}
This form of equations immediately suggests yet another change of
variables

\begin{equation}
\eta=\int_{q}^{w}\frac{dq}{q}e^{-q^{2}/2},\;\;\;\;\Psi=\psi/\rho_{0}\label{eq:eta-def}
\end{equation}
The variable $\eta$ runs from its minimum value, $\eta_{m}<0$ (where
the flow accelerates in $z$- direction to some maximum speed $q_{max}>w$,
where $z\gg a$) to infinity (flow stagnation point, $q\to0).$ In
these new variables the above equations constitute the familiar Cauchy-Riemann
relations for $\chi$ and $\Psi$:

\[
\frac{\partial\chi}{\partial\eta}=-\frac{\partial\Psi}{\partial\vartheta},\;\;\;\frac{\partial\Psi}{\partial\eta}=\frac{\partial\chi}{\partial\vartheta}
\]
Therefore, we can apply powerful tools of complex analysis to our
further derivation of the analytic form of the flow, in a close similarity
to the analyses of the special types of planar flows that are both
irrotational and incompressible. Note, however, that being arguably
irrotational our flow is highly compressible in physical variables.
It is only the Chaplygin form, combined with the above change of variables,
that makes the flow to appear incompressible in the new variables. 

From this point on, we follow the standard methods of complex analysis
and introduce a complex flow potential

\[
\Lambda=\chi+i\Psi
\]
defined on a complex plane

\[
\sigma=\eta+i\vartheta
\]
According to the above Cauchy-Riemann relations, $\Lambda$ is an
analytic function of $\sigma$, and its imaginary and real parts are
harmonic functions

\begin{equation}
\Delta\chi=\Delta\Psi=0,\;\;\;{\rm where\;\;\;\Delta\equiv\frac{\partial^{2}}{\partial\eta^{2}}+\frac{\partial^{2}}{\partial\vartheta^{2}}}\label{eq:LaplaceEgChiPsi}
\end{equation}
Thus, our task is indeed significantly simplified and can be formulated
now as the following Dirichlet problem: find an analytic function
$\Lambda$ in some region on the plane $\sigma$ taking specific value
at its boundary. We therefore need to define this boundary which we
do in two steps. First, we map the flow region in $x,z$ variables
to that in $v_{x},v_{z}$ variables. These regions are shown in the
left and middle panels of Fig.\ref{fig:Hodograph-transform:-physical}. 

To proceed, we break down the boundary of the part of the original
physical flow under consideration in the five segments numbered from
one to five in Fig\ref{fig:Hodograph-transform:-physical}. The segment
1 starts from the shock surface, going along the symmetry axis $z=0$
where the flow decelerates to reach the stagnation point at $x=z=0$.
The flow turns sharply by ninety degrees at this point. So, the segment
2 of the flow boundary goes along another symmetry axis, the $z-$
axis. In reality, the flow continues to infinity along it, gradually
increasing its speed, but we do not follow it beyond some velocity,
$v=q_{m}\lesssim1$, as we pointed out earlier. This rather technical
condition allows for the subsonic approximation throughout the flow
region. We therefore draw the segment 3 connecting $z-$ axis with
the shock surface along the line with the constant $q=q_{m}$. The
remaining segments 4 and 5 go along the shock surface. They are different
in that the upstream ions cross the segment 5 and get reflected from
segment 4, because of the insufficient velocity component normal to
the shock to pass above the shock potential. This distinction is not
easy to make beforehand, as the shock potential decreases with $z$,
Sec.\ref{subsec:Shock-geometry-and}. Moreover, the particle incidence
angle increases, thus showing an opposite trends with regard to the
shock reflectivity. The flow is different from conventional hydrodynamic
flows primarily because it allows for a multi-stream state upstream
owing to particle reflection. On the downstream side of the segment
4, there is no plasma inflowing across the shock and the flow is purely
tangential for the incidence angles $\alpha>\alpha_{cr}$, Fig.\ref{fig:Hodograph-transform:-physical}.

Turning to the Dirichlet boundary conditions for the stream function
$\Psi$, it must remain constant on those segments of the boundary
where the flow is tangential. These are the segments 1,2, and 4. Obviously,
the stream function value on the segments $1$ and 2 are equal and
we can set them to be zero, $\Psi_{1}=\Psi_{2}=0$. Let the respective
constant value on the segment 4 be $\Psi_{4}\neq0$. The difference
between these two values of the stream function equals to the net
mass flux starting from the segments 5 (inflow) and ending at segment
3 (outflow), according to the definition of $\Psi$, given in eq.(\ref{eq:PsiDef}).
Since this is the same flux emitted from the wire into the angle $\alpha_{cr}$,
we find

\begin{equation}
\Psi_{4}=\frac{\alpha_{cr}}{2\pi}S\label{eq:psi4}
\end{equation}
where we have denoted the mass flux per unit length of each wire by
$S$. In the next subsection we consider the Dirichlet problem including
its boundary conditions in more detail. 

\subsection{Dirichlet Problem for the Stream Function and its Solution\label{subsec:Dirichlet-Problem-for}}

The above-discussed boundary conditions relate to the physical plane
of the flow, i.e., $x,z-$ plane. The equations in these variables
are strongly nonlinear and difficult so solve, as discussed in Sec.\ref{subsec:Analysis-of-Colliding}.
By contrast, the ``solvable'' version of those equations given by
eq.(\ref{eq:LaplaceEgChiPsi}) is defined on the hodograph plane $\eta,\vartheta$.
Recall that we have transformed to these variables in two steps, Fig.\ref{fig:Hodograph-transform:-physical}.
First, we mapped $x,z-$ plane to $v_{x},v_{z}$ and then to $\eta,\vartheta$.
The new boundary conditions for the Dirichlet problem merit more attention.
Let us start with considering a counter-clockwise contour formed by
the segments 1-5 on the physical plane $x,z$. The image of this contour
on the $v_{x},v_{z}-$plane is depicted in the middle panel of Fig.\ref{fig:Hodograph-transform:-physical}.
Next, we give a complete formulation of the Dirichlet problem and
describe the transformation of the above contour to the $\eta,\vartheta$
plane, shown in the right panel of Fig.\ref{fig:Hodograph-transform:-physical}. 

In the new variables $\eta,\vartheta$ the Dirichlet problem for $\Psi\left(\eta,\vartheta\right)$
can be formulated in the following way: find a harmonic function $\Psi$, 

\begin{equation}
\frac{\partial^{2}\Psi}{\partial\eta^{2}}+\frac{\partial^{2}\Psi}{\partial\vartheta^{2}}=0,\label{eq:LaplaceForPsi}
\end{equation}
inside a boundary shown in the right panel of Fig.\ref{fig:Hodograph-transform:-physical}
that takes the following values on this boundary:
\begin{enumerate}
\item $\Psi=0$ on segments 1 and 2, 
\item $\Psi=\Psi_{3}\left(\vartheta\right)$ on segment 3
\item $\Psi=\Psi_{4}=const$ on segment 4
\item $\Psi=\Psi_{5}\left(\vartheta\right)$ on segment 5,
\end{enumerate}
where the constant value $\Psi_{4}$ is given by eq.(\ref{eq:psi4}).
The function $\Psi_{5}$$\left(\vartheta\right)$ monotonically increases
from $0$ to $\Psi_{4}$ in the interval $0\le\vartheta\le\pi/2-\vartheta_{s}$
and the function $\Psi_{3}\left(\vartheta\right)$ monotonically decreases
from $\Psi_{4}$ to 0, in the interval $\pi/2-\vartheta_{s}<\vartheta\le\pi/2$.

As we already mentioned, it is the boundary conditions that make the
Dirichlet problem difficult to solve, not the eq.(\ref{eq:LaplaceForPsi}).
Therefore, the common strategy for solving such problems is in mapping
the problem domain to a simpler one, using a conformal map to a new
complex variable $\zeta$, that is $\eta+i\vartheta=\sigma\to\zeta$.
As the map is conformal, it does not change the Laplace equation for
$\Psi$ that remains analytic also in the new variables. The semi-infinite
domain in the $\eta,\sigma$- plane shown on the right panel of Fig.\ref{fig:Hodograph-transform:-physical}
can be viewed as either a hexagon with two right angles at $\eta=\infty$,
$\sigma=0,\pi$, or a pentagon with just one zero angle at $\eta=\infty$.
We wish to map the region confined by the segments 1-5 on the $\sigma-$
plane to the upper half of $\zeta-$ plane, so the boundary on the
$\sigma-$ plane must be mapped to the real axis of the $\zeta-$
plane. This map can most efficiently be built by using the Schwartz-Christoffel
integral. From its perspective, the domain on the $\sigma-$ plane
is to be regarded as a pentagon with one zero angle at the infinity,
$\eta=\infty$, rather than a hexagon also mentioned above. We lay
aside this latter interpretation of the semi-infinite domain until
Sec.\ref{subsec:Simple-Flow-Example}, where we consider a simplified
flow example.

Here, we establish the correspondence between the five points on the
real axis of the new $\zeta-$ plane $\zeta_{3}<\zeta_{2}<\zeta_{1}<0,1$,
and the angles of the pentagon on $\sigma-$ plane. We will distinguish,
however, between the points $\eta=\infty$, $\vartheta=0,i\pi$ on
the $\sigma-$ plane by ascribing different arguments to the linear
function $1-\zeta$, involved in the Schwartz-Christoffel integral.
The correspondence is as follows:
\begin{enumerate}
\item $\sigma=0$ $\to$ $\zeta=0$
\item $\sigma=i\left(\pi/2-\vartheta_{s}\right)$ $\to\zeta=\zeta_{1}$\label{enu:BCZet1}
\item $\sigma=\eta_{m}+i\left(\pi/2-\vartheta_{s}\right)$ $\to\zeta=\zeta_{2}$\label{enu:BCzet2}
\item $\sigma=\eta_{m}+i\pi/2$ $\to\zeta=\zeta_{3}$\label{enu:BCzet3}
\item $\sigma=\infty+i0$ $\to\zeta=1,\;\arg\left(\zeta-1\right)=\pi$ 
\item $\sigma=\infty+i\pi/2$ $\to\zeta=1,\;\arg\left(\zeta-1\right)=0$
\end{enumerate}
Then, the Schwartz-Christoffel map of the real axis $\zeta$ to the
flow region on the $\sigma$ plane can be written as follows:

\begin{equation}
\sigma=\Theta\int_{0}^{\zeta}\sqrt{\frac{\zeta-\zeta_{1}}{\left(\zeta-\zeta_{2}\right)\left(\zeta-\zeta_{3}\right)}}\frac{d\zeta}{\left(1-\zeta\right)\sqrt{\zeta}}\label{eq:SchwChristFull}
\end{equation}
Some details of the derivation of this result can be found in Appendix
\ref{sec:Schwartz-Christoffel-transform}. Now writing the new variable
$\zeta$ as $\zeta=\zeta^{\prime}+i\zeta^{\prime\prime}$, the solution
of the Dirichlet problem for the stream function can be represented
in the form of Poisson integral

\begin{equation}
\Psi\left(\zeta^{\prime},\zeta^{\prime\prime}\right)=\frac{\zeta^{\prime\prime}}{\pi}\int_{-\infty}^{\infty}\frac{\Psi\left(t,0\right)dt}{\left(t-\zeta^{\prime}\right)^{2}+\zeta^{\prime\prime2}}\label{eq:DirichletSol}
\end{equation}
The earlier formulated boundary conditions for the Dirichlet problem
require the function $\Psi\left(\zeta^{\prime},0\right)$ under the
integral to be defined as follows

\begin{equation}
\Psi\left(\zeta^{\prime},0\right)=\begin{cases}
0, & -\infty<\zeta^{\prime}\le\zeta_{3}\\
\Psi_{3}\left(\zeta^{\prime}\right), & \zeta_{3}<\zeta^{\prime}<\zeta_{2}\\
\Psi_{4}=const, & \zeta_{2}\le\zeta^{\prime}\le\zeta_{1}\\
\Psi_{5}\left(\zeta^{\prime}\right), & \zeta_{1}<\zeta^{\prime}<0\\
0, & 0\le\zeta^{\prime}<\infty
\end{cases}\label{eq:Psi5cases}
\end{equation}
where $\Psi_{4}=const$ is given by eq.(\ref{eq:psi4}). The boundary
functions $\Psi_{5}\left(\zeta\right)$ can be defined by considering
the flow in front of the shock. Using the cylindrical coordinates
for velocity, by the definition of the stream function, eq.(\ref{eq:PsiDef}),
we have

\[
r\rho w_{0}=\frac{\partial\Psi}{\partial\alpha}=\frac{S}{2\pi},
\]
(cf. eq.{[}\ref{eq:psi4}{]}), where $r$ is the distance to the wire;
according to the notations of Sec. \pageref{subsec:Shock-geometry-and},
we denote the angle made by the flow velocity with the $x-$ axis
upstream by $\alpha$. Here $w_{0}$ is the radial velocity of the
flow coming from the wire center at $x=-a$. Therefore, ahead of the
shock the stream function changes linearly with $\alpha$

\begin{equation}
\Psi=\frac{\alpha}{2\pi}S\label{eq:psi5ofalfa}
\end{equation}
Now we need to relate the angle $\alpha$ upstream with that of downstream
of the shock, $\vartheta$:
\[
\cos\vartheta=\frac{v_{n}\cos\beta-v_{\tau}\sin\beta}{\sqrt{w_{0}^{2}-2\phi_{s}}}
\]
where $v_{n}$ and $v_{\tau}$ are the normal and tangential to the
local shock normal components of the flow velocity downstream, $\beta$
is the shock inclination angle, and $\phi_{s}$ is the cross-shock
potential (see eqs.{[}\ref{Vtau}-\ref{eq:Vn}{]}). Note that in our
construction of the hodograph plane in Sec.\ref{subsec:Hodograph-Equations},
the shock potential $\phi_{s}\approx const$ for $\alpha<\alpha_{cr}$,
Fig.\ref{fig:Hodograph-transform:-physical}. At this critical angle
the incoming ions begin to reflect off the shock. If we were to relax
this simplifying condition, the flow velocity on the sector $5$ would
not be constant, thus making the Dirichlet problem more difficult
to solve. For $\alpha>\alpha_{cr}$ the dependence of $\phi_{s}$
on $z$ is essential and we have given an approximate expression for
it in eq.(\ref{eq:fi-s-ofz}). We also argued for the smallness of
shock inclination angle, $\beta\ll1$, which simplifies the last expression
to the following one

\[
\cos\vartheta=\frac{\sqrt{w_{0}^{2}\cos^{2}\alpha-2\phi_{s}}}{\sqrt{w_{0}^{2}-2\phi_{s}}}
\]
Note that $\beta\ll1$ condition also provides more justification
for the constant $\phi_{s}$ approximation at $\alpha<\alpha_{cr}$.
By denoting $U=2\phi_{s}/w_{0}^{2}$, we can rewrite the last expression 

\[
\cos\vartheta=\frac{\sqrt{\cos^{2}\alpha-U}}{\sqrt{1-U}}=\sqrt{1-\frac{\sin^{2}\alpha}{1-U}}
\]
which relates $\alpha$ and $\vartheta$ in a simple way: 

\[
\sin\alpha=\sqrt{1-U}\sin\vartheta
\]
Considering the case of strong shock reflectivity, $1-U\ll1$, we
see that particle reflection begins at small $\alpha$, since $\sin\alpha_{cr}=\sqrt{1-U}\ll1.$
Therefore, we can indeed treat the cross-shock potential as being
constant at $\alpha<\alpha_{cr},$ that is $\phi_{s}\approx\phi_{s}\left(\alpha=0\right)=const$.
However, $\phi_{s}$ significantly changes in the outflow region on
segment 4, above the critical point at $\alpha_{cr}$, Fig.\ref{fig:Hodograph-transform:-physical}.
Using the last relation we can specify the boundary function $\Psi_{5}\left(\vartheta\right)$,
introduced in Sec.\ref{subsec:Dirichlet-Problem-for}. Indeed for
the case of $U\approx1,$we have 

\[
\Psi_{5}\left(\vartheta\right)=\frac{S}{2\pi}\sqrt{1-U}\sin\vartheta=\Psi_{4}\sin\vartheta,
\]
 in accordance with eq.(\ref{eq:psi4}).

The remaining element of the boundary function in the solution $\Psi\left(\zeta^{\prime},0\right)$
of the Dirichlet problem given in eq.(\ref{eq:DirichletSol}) is the
outflow stream function $\Psi_{3}$. Recall that we have chosen this
boundary from a mere requirement that the flow is significantly accelerated
toward this region as to make the electrostatic potential decreased
significantly below its peak at the stagnation point $x=z=0$ (see
eq.{[}\ref{eq:Bern2}{]}). Since no particles can enter the flow from
the shock surface in the segment 3 region, our assumption about the
small shock inclination angle, $\vartheta_{s}\ll1$, holds up. This
means that the flow is well aligned with $z-$ axis, so $v_{x}\ll v_{z}$.
As the flow is irrotational, $\partial v_{z}/\partial x\approx0$,
so we can approximate the stream function by a linear function in
$x$, or $\vartheta$:

\[
\Psi_{3}\left(\vartheta\right)=\frac{\pi/2-\vartheta}{\pi/2-\vartheta_{s}}\Psi_{4},\;\;\;{\rm for}\;\;\;\vartheta_{s}\le\vartheta\le\pi/2
\]
This concludes our calculation of the boundary function $\Psi\left(\sigma\right)$
for the Dirichlet solution, given in eq.(\ref{eq:DirichletSol}).
Two simple functions of the velocity angle $\vartheta$, $\Psi_{3}$
and $\Psi_{5}$, obtained above, \foreignlanguage{american}{essentialize}
the $\Psi\left(\sigma\right)$ variation along the boundary, while
$\Psi\left(\sigma\right)$ takes constant values on the remaining
boundary segments 1,2 and 4. Expressing $\vartheta$ in $\Psi_{3}$
and $\Psi_{5}$ through $\zeta^{\prime}$, using eq.(\ref{eq:SchwChristFull}),
and substituting the result into eq.(\ref{eq:DirichletSol}), provide
a complete solution for the stream function inside the region of shocked
plasma. More difficult, of course, is to convert the $\vartheta-$
dependence of $\Psi_{3}$ and $\Psi_{5}$ into $\zeta-$ dependence
and back, as such conversion requires the determination of the Schwartz-Christoffel
parameters, as discussed in Appendix \ref{sec:Schwartz-Christoffel-transform}.
We defer this task to a future study, as it cannot be completed using
only the analytic solutions, to which we adhere in this paper. However,
we present below a simplified version of such calculation that can
still be completed analytically. 

\subsection{Simple Flow Example\label{subsec:Simple-Flow-Example} }

In the preceding subsection, we obtained a general solution for the
stream function of the flow between two termination shocks in form
of the Poisson integral given by eq.(\ref{eq:DirichletSol}). The
solution depends on the boundary function $\Psi$, also specified
in Sec.\ref{subsec:Dirichlet-Problem-for}. The problem of transforming
this result back to the variable $\sigma$, that is $\zeta\mapsto\sigma$,
is not simple, given the complexity of the relation between these
two variables shown in eq.(\ref{eq:SchwChristFull}). In particular,
to determine the four parameters of this transform, $\Theta,\zeta_{1-3}$,
we need to solve four transcendental equations, eqs.(\ref{eq:SCparam1}-\ref{eq:SCparam4}).
At the same time, the relation between $\zeta$ and $\sigma$ variables,
established by the Schwartz-Christoffel transform, can be significantly
simplified by neglecting some details of the post shock flow. The
rationale behind such simplification is as follows.

The boundary function $\Psi$ rises from zero to its maximum value
$\Psi=\Psi_{4}$ (reached on segment 4) and back to zero while the
contour runs along the segments 3,4, and 5. Let us replace these three
connected segments by a single line connecting points $\sigma=\eta_{m}+i\pi/2$
and $\sigma=0$. By the principle of analytic continuation, the Poisson
integral in eq.(\ref{eq:DirichletSol}) will give the same solution
inside of the new boundary provided that we know the value of the
stream function on this new boundary. Then, we can map it to the appropriate
interval on the $\zeta-$ plane to replace the part of the integral
in eq.(\ref{eq:DirichletSol}) between $\zeta=\zeta_{3}$ and $\zeta=0$.
It is clear that the conformal map $\sigma\mapsto\zeta$ becomes much
simpler since the three sides (segments 3-5) replaced by one line
specified above. Let us denote by $\pi\alpha_{*}$ the angle this
line makes with the abscissa on $\sigma-$ plane:

\[
\tan\pi\alpha_{*}=-\frac{\pi}{2\eta_{m}}>0
\]
The new conformal map can be written as follows

\[
\sigma=\frac{1}{2}\int_{0}^{\zeta}\zeta^{-\alpha_{*}}\left(1+\zeta\right)^{\alpha_{*}-1}d\zeta
\]
It transforms the semi-strip on $\sigma-$ plane formed by the line
$\eta_{m}+i\pi/2,\;0$ and two semi-infinite lines, $0,+\infty$;
$\eta_{m}+i\pi/2,\;\infty+i\pi/2$ to the upper half-plane of complex
$\zeta$. Even better sense of this simplified map one can get by
considering the case of small $\left|\eta_{m}\right|\ll1$. As $\alpha_{*}\approx1/2$
in this case the map is simply $\zeta=\sinh^{2}\sigma$ which is easy
to check by direct substitution of this relation into the above integral.
By rewriting this map in the expanded form

\[
\zeta^{\prime}+i\zeta^{\prime\prime}=\frac{1}{2}\cosh2\eta\;\cos2\vartheta-\frac{1}{2}+\frac{i}{2}\sinh2\eta\;\sin2\vartheta,
\]
we can change variables in the Poisson integral in eq.(\ref{eq:DirichletSol})
and express the solution through $\vartheta,\eta$ which are directly
related to the flow velocity. We also make use of the boundary function
$\Psi\left(\zeta,0\right)$ being non-zero only in the interval $-1<\zeta<0$.
This interval corresponds to the $0<\vartheta<\pi/2$ interval on
the flow velocity plane. The Poisson integral then rewrites

\[
\Psi\left(\eta,\vartheta\right)=\frac{2}{\pi}\sinh2\eta\;\sin2\vartheta\int_{0}^{\pi/2}\frac{\sin2\vartheta^{\prime}\;\Psi\left(\vartheta^{\prime},0\right)d\vartheta^{\prime}}{\left(\cosh2\eta\;\cos2\vartheta-\cos2\vartheta^{\prime}\right)^{2}+\sinh^{2}2\eta\;\sin^{2}2\vartheta}
\]
While being grossly simplified, this expression is helpful for understanding
the flow structure in the shocked plasma region. The variable $\eta$
can be expressed directly through the local velocity $q$ using eq.(\ref{eq:eta-def}).
The formulas given in Appendix \ref{sec:Hodograph-Transform-App}
allows one to map the velocity plane to the physical plane $x,z$.
Using Bernoulli integral, eq.(\ref{eq:Bern2}), the flow velocity
dependence on $x,z$ explicitly provides the distribution of the electrostatic
potential on the physical plane which is a key ingredient in calculating
the growth rate obtained in Sec.\pageref{subsec:Instability-Analysis}. 

\section{Summary and Outlook\label{sec:Conclusions-and-Outlook}}

In this paper we studied colliding plasmas concentrically flowing
from two ohmically exploded wires. Assuming that the flows are mildly
supersonic, our main interest was centered around a thin layer of
shocked plasma that forms between the colliding flows. We focused
on two properties of the layer that can be categorized as microscopic
and macroscopic. The microscopics concerns instabilities developing
at scales much smaller than the layer itself. Given the anticipated
macroscopic characteristics of this plasma, the main candidate is
the lower-hybrid drift instability (LHDI), primarily powered by the
ExB electron drift. The magnetic field, supporting the drift, is generated
at earlier times in the outflown plasmas by currents driven through
the wires during their explosion. The ambipolar electric field, which
is also needed to produce the drift, emerges from the collision process.
This field is also one of the key elements of the macroscopic description
of the shocked plasma flow. 

Generally, the investigated collision of magnetized plasmas encompasses
several elements other than the macroscopic electric field, which
are difficult to separate from one another. Before the LHDI sets on,
two shocks form that terminate each flow. Oppositely directed magnetic
fields are carried toward each other by shocked plasmas and they will
eventually reconnect near the mid-plane between the shocks. These
processes need to be studied self-consistently, including the LHDI
while it provides the anomalous resistivity that, in turn, determines
the field reconnection rate in low-collisionality regimes. It is clear,
however, that all these phenomena are difficult for studying them
self-consistently with mere analytic methods. Therefore, in this paper
we addressed the instability and the flow past the shocks independently.
This task separation is tenable because we have found the solution
of the linear dispersion relation for the instability without imposing
specific restrictions on the characteristics of the flow and unstable
waves. Such restrictions usually follow from the analysis of macroscopic
equilibrium. 

For the above-mentioned reasons, our study of the macroscopics of
the shocked plasma flow is semi-independent of the instability study.
More importantly, the macroscopic consequences of the instability
can be included in a form of pondermotive pressure exerted by Lower-hybrid
waves on the shocked plasma flow. The main results obtained at this
phase of research are the following:
\begin{itemize}
\item the growth rate of the lower-hybrid drift instability in the post-shock
plasma flow between two exploding wires is calculated in a closed
form without further simplifications, customarily applied to similar
instability analyses
\item the plasma collision process is characterized including the determination
of the stand-off distances of the termination shocks formed in the
respective flows and the profile of the macroscopic electric field
between them
\item Chaplygin equations for the flow past the termination shocks are derived
using the hodograph map
\item general analytic solution for the stream function of the flow is obtained
using the Schwartz-Christoffel map of the hodograph plane
\end{itemize}
These results lay the ground for a more complete characterization
of the collision of magnetized plasmas resulting from the wire explosion
or otherwise. As the next logical steps they suggest:
\begin{itemize}
\item to study macroscopic implications of the LHDI by calculating the electron
heating and anomalous resistivity
\item to investigate the impact of pondermotive pressure on the flow which
was shown to strongly depends on $T_{e}/T_{i}$ ratio
\item reiterated consideration of the shocked plasma flow with the LHDI
effects included
\item study of the magnetic field reconnection in the shocked plasma based
on the anomalous resistivity supported by the LHDI
\item study of the generation of mesoscale structures resulting from LHDI,
ExB shear flow, and field reconnection
\end{itemize}
The last item is worth commenting briefly as it embodies a wide range
of interrelated processes of experimental significance. The importance
of the mesoscale structures \textendash{} which are larger than the
lower-hybrid wave length but smaller than the plasma layer \textendash{}
is that they can be probed and visualized by the available laser diagnostics
directly \cite{Sarkisov2005}. We are planning to explore a number
of mechanisms for their generation. These include a spectral transformations
to the mesoscale structures from short scale lower-hybrid waves via
interactions with acoustic-type perturbations, primarily based on
the modulational instability \cite{Sotnikov1978FizPl}. Other candidates
for their formation are the shear flow generated vortices \cite{Sotnikov2010ITPS},
as well as tearing and modulational instabilities in the neutral sheet
area \cite{1985SvJPP..11.1096M}.

\acknowledgments

This work was supported by the Air Force Office of Scientific Research under grant LRIR \# 16RYCOR289.
Public release approval record: 88ABW-2018-4241

\appendix

\section{Hodograph Transform\label{sec:Hodograph-Transform-App}}

We use that particular type of hodograph transform in which the role
of the new dependent variables is assumed by the flow potential and
the stream function, eqs.(\ref{eq:chiDef}) and (\ref{eq:PsiDef}),
while the role of the independent variables is left for the velocity
components. By definition, the flow potential $\chi$ and stream function
$\psi$ obey the following relations

\begin{eqnarray*}
d\chi & = & v_{x}dx+v_{z}dz\\
d\psi & = & -v_{z}\rho dx+v_{x}\rho dz
\end{eqnarray*}
from which we obtain

\begin{eqnarray*}
dx & = & \frac{\cos\vartheta}{q}d\chi-\frac{\sin\vartheta}{\rho q}d\psi\\
dz & = & \frac{\cos\vartheta}{\rho q}d\psi+\frac{\sin\vartheta}{q}d\chi
\end{eqnarray*}
Here and below we use the cylindrical coordinates in the velocity
space $q$ and $\vartheta$ introduced in eq.(\ref{eq:VelPolar}).
The last two equations manifest $dx$ and $dz$ as complete differentials
of functions $x\left(\chi,\psi\right)$ and $z\left(\chi,\psi\right)$.
The respective cross-derivatives of the coefficients at $d\chi$ and
$d\Psi$ must then be equal:

\begin{eqnarray*}
\frac{\partial}{\partial\psi}\frac{\cos\vartheta}{q} & = & -\frac{\partial}{\partial\chi}\frac{\sin\vartheta}{q\rho}\\
\frac{\partial}{\partial\psi}\frac{\sin\vartheta}{q} & = & \frac{\partial}{\partial\chi}\frac{\cos\vartheta}{q\rho}
\end{eqnarray*}
This system of equations can be rewritten as follows

\begin{eqnarray*}
\left[\frac{\partial}{\partial\chi}\frac{1}{q\rho}-\frac{1}{q}\frac{\partial\vartheta}{\partial\psi}\right]\sin\vartheta+\left[\frac{\partial}{\partial\psi}\frac{1}{q}+\frac{1}{q\rho}\frac{\partial\vartheta}{\partial\chi}\right]\cos\vartheta & = & 0\\
\left[\frac{\partial}{\partial\chi}\frac{1}{q\rho}-\frac{1}{q}\frac{\partial\vartheta}{\partial\psi}\right]\cos\vartheta-\left[\frac{\partial}{\partial\psi}\frac{1}{q}+\frac{1}{q\rho}\frac{\partial\vartheta}{\partial\chi}\right]\sin\vartheta & = & 0
\end{eqnarray*}
The terms in the brackets may be regarded as two unknown entering
a linear system of two equations. Since the determinant of this system
is ${\rm sin^{2}\vartheta+\cos^{2}\vartheta=1},$ it has only a trivial
solution. In other words, each term in the brackets must be zero.
Substituting $\rho$ from eq.(\ref{eq:rhoOfqDef}), we write these
two conditions as follows:

\begin{eqnarray*}
\frac{\partial\vartheta}{\partial\psi} & = & \frac{q^{2}-1}{\rho q}\frac{\partial q}{\partial\chi}\\
\frac{\partial\vartheta}{\partial\chi} & = & \frac{\rho}{q}\frac{\partial q}{\partial\psi}
\end{eqnarray*}
By swapping the dependent and independent variables in these equations,
we obtain Chaplygin eqs.(\ref{eq:Chapl1}-\ref{eq:Chapl2}). The inversion
of the derivatives in these equations can be achieved by writing identities
of the following kind

\[
\frac{\partial\vartheta}{\partial\psi}=\frac{\partial\left(\vartheta,\chi\right)}{\partial\left(\psi,\chi\right)}=\frac{\partial\left(\vartheta,q\right)}{\partial\left(\psi,\chi\right)}\frac{\partial\left(\vartheta,\chi\right)}{\partial\left(\vartheta,q\right)}=J\frac{\partial\chi}{\partial q}
\]
where the Jacobian of the transform between the two sets of variables
is
\[
J=\frac{\partial\left(\vartheta,q\right)}{\partial\left(\psi,\chi\right)}.
\]

\section{Schwartz-Christoffel transform\label{sec:Schwartz-Christoffel-transform}}

The Schwartz-Christoffel transform maps a polygon on a complex plane
to the upper half of a new plane. In our case we need to map the pentagon
on the $\sigma=\eta+i\vartheta-$ plane shown in Fig.\ref{fig:Hodograph-transform:-physical}
(right panel) to the upper half-plane of the complex plane $\zeta$.
The four right angles of the pentagon at the finite points of $\sigma-$
plane will correspond to four finite points of the real $\zeta-$
axis: $\zeta_{3}<\zeta_{2}<\zeta_{1}<0$, while the two points at
$\eta=\infty$ are considered as one zero-degree angle of the pentagon,
as indicated in Sec.\ref{subsec:Dirichlet-Problem-for}. As we know
from the complex analysis, an integral of the kind

\begin{equation}
\sigma=\Theta\int_{\zeta_{0}}^{\zeta}\left(\zeta-\zeta_{1}\right)^{\alpha_{1}/\pi-1}\dots\left(\zeta-\zeta_{n}\right)^{\alpha_{n}/\pi-1}\label{eq:SC1}
\end{equation}
maps a polygon with $n$ angles on the $\sigma-$ plane to the upper
half-plane $\eta$. Here the constants $\alpha_{k}$ are the internal
angles of the polygon. According to the point-to-point correspondence
list (1-6) in Sec.\ref{subsec:Dirichlet-Problem-for}, we can specify
the anchor points $\zeta_{k}$ and angles $\alpha_{k}$ to obtain
the expression for $\sigma$ shown in eq.(\ref{eq:SchwChristFull}):

\[
\sigma=\Theta\int_{0}^{\zeta}\sqrt{\frac{\zeta-\zeta_{1}}{\left(\zeta-\zeta_{2}\right)\left(\zeta-\zeta_{3}\right)}}\frac{d\zeta}{\left(1-\zeta\right)\sqrt{\zeta}}
\]
Apart from preserving the origin ($1/\sqrt{\zeta}$ factor in the
integrand), this choice of the transform parameters $\zeta_{k},\alpha_{k}$
connects, in particular, the $\zeta=1$ pole to the zero-angle point
of the pentagon on the right panel of Fig.\ref{fig:Hodograph-transform:-physical}.
The remaining $\zeta_{1-3}$ anchor points correspond to the two $\pi/2$
angles and one $3\pi/2$ angle of the pentagon. The integral thus
maps the interval $\zeta_{1}<\zeta<0$ to the interval of the imaginary
$\sigma-$ axis, $0<\Im\sigma<\pi/2-\vartheta_{s}$. The branch of
$\sqrt{\zeta}$ is such that $\sqrt{\zeta}>0$ for $\zeta>0$. The
remaining two finite sides of the pentagon are mapped from the intervals ($\zeta_{2},\zeta_{1}$)
and ($\zeta_{3},\zeta_{2})$. The two semi-infinite sides of the pentagon
are related to the point $\zeta=1$ with $\arg\left(1-\zeta\right)=0,\pi$.
The transition from the lower to upper branch occurs when the contour
on $\zeta-$ plane goes over the $\zeta=1$ pole. As the $\Im\sigma$
jumps from $0$ to $\pi/2$, we have the following relation for the
parameters of the Schwartz-Christoffel integral:

\begin{equation}
\Theta\sqrt{\frac{1-\zeta_{1}}{\left(1-\zeta_{2}\right)\left(1-\zeta_{3}\right)}}=\frac{1}{2}\label{eq:SCparam1}
\end{equation}
The contour then goes to $\Re\zeta=\infty$, and the closing arc $\left|\zeta\right|=\infty$
is therefore mapped to some point on the upper semi-infinite branch,
yielding no further relation for the parameters of the map. By requiring
that the integrals between the remaining points $\zeta_{1-3}$, including
the origin, produce the appropriate changes in $\sigma$, as listed
under the items (1-6) in Sec.\ref{subsec:Dirichlet-Problem-for},
we find the following relations

\begin{equation}
\vartheta_{s}=\Theta\int_{\zeta_{3}}^{\zeta_{2}}\sqrt{\frac{\zeta-\zeta_{1}}{\left(\zeta_{2}-\zeta\right)\left(\zeta-\zeta_{3}\right)}}\frac{d\zeta}{\left(1-\zeta\right)\sqrt{\zeta}}\label{eq:SCparam2}
\end{equation}

\begin{equation}
\eta_{m}=-\Theta\int_{\zeta_{2}}^{\zeta_{1}}\sqrt{\frac{\zeta-\zeta_{1}}{\left(\zeta-\zeta_{2}\right)\left(\zeta-\zeta_{3}\right)}}\frac{d\zeta}{\left(1-\zeta\right)\sqrt{\zeta}}\label{eq:SCparam3}
\end{equation}

\begin{equation}
\frac{\pi}{2}-\vartheta_{s}=\Theta\int_{\zeta_{1}}^{0}\sqrt{\frac{\zeta-\zeta_{1}}{\left(\zeta-\zeta_{2}\right)\left(\zeta-\zeta_{3}\right)}}\frac{d\zeta}{\left(1-\zeta\right)\sqrt{-\zeta}}\label{eq:SCparam4}
\end{equation}
Eqs.(\ref{eq:SCparam1}-\ref{eq:SCparam4}) are thus four equations
for four Schwartz-Christoffel parameters $\Theta,\zeta_{1-3}$, given
the flow region boundaries described here by the two values of $\eta_{m}$
and $\vartheta_{s}$.

\bibliographystyle{plain}
\bibliography{MalkovSotnikovAR.bbl}

\begin{thebibliography}{10}

\bibitem{Chittenden1995}
J.~P. {Chittenden}.
\newblock {The effect of lower hybrid instabilities on plasma confinement in
  fiber Z pinches}.
\newblock {\em Physics of Plasmas}, 2:1242--1249, April 1995.

\bibitem{Chittenden2001}
J.~P. {Chittenden}, S.~V. {Lebedev}, S.~N. {Bland}, F.~N. {Beg}, and M.~G.
  {Haines}.
\newblock {One-, two-, and three-dimensional modeling of the different phases
  of wire array Z-pinch evolution}.
\newblock {\em Physics of Plasmas}, 8:2305--2314, May 2001.

\bibitem{DavidsonGladd1975}
R.~C. {Davidson} and N.~T. {Gladd}.
\newblock {Anomalous transport properties associated with the
  lower-hybrid-drift instability}.
\newblock {\em Physics of Fluids}, 18:1327--1335, October 1975.

\bibitem{Dieckmann2018PPCF...60a4014D}
M.~E. {Dieckmann}, D.~{Doria}, G.~{Sarri}, L.~{Romagnani}, H.~{Ahmed},
  D.~{Folini}, R.~{Walder}, A.~{Bret}, and M.~{Borghesi}.
\newblock {Electrostatic shock waves in the laboratory and astrophysics:
  similarities and differences}.
\newblock {\em Plasma Physics and Controlled Fusion}, 60(1):014014, January
  2018.

\bibitem{Drake2018}
R.P. Drake.
\newblock {\em High Energy Density Physics}.
\newblock Graduate Texts in Physics Series. Springer-Verlag, 2018.

\bibitem{Gurevich68}
A.~V. {Gurevich}.
\newblock {Distribution of Captured Particles in a Potential Well in the
  Absence of Collisions}.
\newblock {\em Soviet Journal of Experimental and Theoretical Physics}, 26:575,
  March 1968.

\bibitem{Kevlahan97}
{N.~K.-R.} {Kevlahan}.
\newblock {The vorticity jump across a shock in a non-uniform flow}.
\newblock {\em Journal of Fluid Mechanics}, 341:371--384, June 1997.

\bibitem{LLFM}
L.~D. {Landau} and E.~M. {Lifshitz}.
\newblock {\em {Fluid Mechanics}}.
\newblock Pergamon Press, 1987.

\bibitem{LiewerKrall1973}
P.~C. {Liewer} and N.~A. {Krall}.
\newblock {Self-consistent approach to anomalous resistivity applied to theta
  pinch experiments}.
\newblock {\em Physics of Fluids}, 16:1953--1963, November 1973.

\bibitem{MSetal_IAshocks2016}
M.~A. Malkov, R.~Z. Sagdeev, G.~I. Dudnikova, T.~V. Liseykina, P.~H. Diamond,
  K.~Papadopoulos, C.-S. Liu, and J.~J. Su.
\newblock Ion-acoustic shocks with self-regulated ion reflection and
  acceleration.
\newblock {\em Physics of Plasmas}, 23(4), 2016.

\bibitem{1985SvJPP..11.1096M}
M.~A. {Malkov} and V.~I. {Sotnikov}.
\newblock {Lower hybrid drift instability and reconnection of magnetic lines of
  force}.
\newblock {\em Soviet Journal of Plasma Physics}, 11:1096--1105, \#sep\# 1985.

\bibitem{MoisSagd63}
S.~S. {Moiseev} and R.~Z. {Sagdeev}.
\newblock {Collisionless shock waves in a plasma in a weak magnetic field}.
\newblock {\em Journal of Nuclear Energy}, 5:43--47, January 1963.

\bibitem{Note1}
Note that in the case of the equivalent colliding flow configuration, the role
  of reflected ions is taken by the ions from the opposite flow.

\bibitem{Ryutov2013PhPlreconnection}
D.~D. {Ryutov}, N.~L. {Kugland}, M.~C. {Levy}, C.~{Plechaty}, J.~S. {Ross}, and
  H.~S. {Park}.
\newblock {Magnetic field advection in two interpenetrating plasma streams}.
\newblock {\em Physics of Plasmas}, 20(3):032703, March 2013.

\bibitem{Sagdeev66}
R.~Z. {Sagdeev}.
\newblock {Cooperative Phenomena and Shock Waves in Collisionless Plasmas}.
\newblock {\em Reviews of Plasma Physics}, 4:23--+, 1966.

\bibitem{Sarkisov2005}
G.~S. {Sarkisov}, S.~E. {Rosenthal}, K.~R. {Cochrane}, K.~W. {Struve},
  C.~{Deeney}, and D.~H. {McDaniel}.
\newblock {Nanosecond electrical explosion of thin aluminum wires in a vacuum:
  Experimental and computational investigations}.
\newblock {\em \pre}, 71(4):046404, April 2005.

\bibitem{Sotnikov2010ITPS}
V.~I. {Sotnikov}, J.~N. {Leboeuf}, and S.~{Mudaliar}.
\newblock {Scattering of Electromagnetic Waves in the Presence of Wave
  Turbulence Excited by a Flow With Velocity Shear}.
\newblock {\em IEEE Transactions on Plasma Science}, 38:2208--2218, September
  2010.

\bibitem{Sotnikov1978FizPl}
V.~I. {Sotnikov}, V.~D. {Shapiro}, and V.~I. {Shevchenko}.
\newblock {Macroscopic consequences of collapse at the lower hybrid resonance}.
\newblock {\em Fizika Plazmy}, 4:450--459, March 1978.

\bibitem{Suttle2018PhPl...25d2108S}
L.~G. {Suttle}, J.~D. {Hare}, S.~V. {Lebedev}, A.~{Ciardi}, N.~F. {Loureiro},
  G.~C. {Burdiak}, J.~P. {Chittenden}, T.~{Clayson}, J.~W.~D. {Halliday},
  N.~{Niasse}, D.~{Russell}, F.~{Suzuki-Vidal}, E.~{Tubman}, T.~{Lane},
  J.~{Ma}, T.~{Robinson}, R.~A. {Smith}, and N.~{Stuart}.
\newblock {Ion heating and magnetic flux pile-up in a magnetic reconnection
  experiment with super-Alfv{\'e}nic plasma inflows}.
\newblock {\em Physics of Plasmas}, 25(4):042108, April 2018.

\end{thebibliography}

\end{document}